\documentclass[endfloats*,prb,preprint,showpacs,preprintnumbers,amsmath,amssymb]{revtex4}

\usepackage{graphicx}
\usepackage{natbib}
\usepackage{dcolumn}
\usepackage{bm}

\begin{document}
  
\newcommand{\kp}{\boldmath$k.p$\unboldmath}
%simple command for use in \section
\newcommand{\kps}{\boldmath$k.p$}

\preprint{APS/123-QED}
  
\title{ Strained Si, Ge and SiGe alloys modeling with full-zone \kp\ method optimized from first principle calculation}
            
\author{D. Rideau} 
 \email{denis.rideau@st.com}
\author{M. Feraille}
 \altaffiliation[Also at ]{Physics Department, INSA Lyon, France.}
\author{L. Ciampolini}
\author{M. Minondo}
\author{C. Tavernier }
\author{H. Jaouen}
\affiliation{
STmicroelectronics\\
rue Jean Monnet, 850\\
BP 16, F-38926 Crolles CEDEX\\
France. 
}
\author{A. Ghetti}
\affiliation{
STmicroelectronics\\
via C. Olivetti, 2\\
I-20041 Agrate Brianza\\
Italy. 
}

\date{\today}
      
\begin{abstract}   
The electronic energy band structure of strained and unstrained Si, Ge and 
SiGe alloys is examined in this work using thirty-level \kp\ analysis.
The energy bands are at first obtained with ab initio calculations based on 
the Local-Density-Approximation of Density-Functional Theory, including 
a GW correction and relativistic effects.  The so-calculated band structure 
is then used to extract the unknown \kp\ fitting parameters with a 
conjugate-gradient optimization procedure. In a similar manner, the results 
of ab initio calculations for strained materials are used to fit the unknown
deformation potentials that are included in the present \kp\ Hamiltonian
following the Pikus and Bir correction scheme.
 We show that the present \kp\ model is an efficient numerical method, as
far as computational time is concerned, that reproduces accurately the
overall band structure, as well as the bulk effective density 
of states and the carrier effective masses, for both strained and unstrained 
materials. As an application, the present thirty-level \kp\ model is used
to describe the band offsets and the variations of the carrier effective 
masses in a strained material, a Si$_{1-x}$Ge$_x$/Si$_{1-y}$Ge$_y$ layer 
system.
\end{abstract}
 
\pacs{71.15.-m, 71.20.Nr, 71.15.Mb, 71.20.-b}
\maketitle 
     
\section{\label{introduction}Introduction}

The continuous downscaling of MOSFET critical dimensions such as the gate length
and the gate oxide thickness has been a very successful trend in current
manufacturing, as testified, e.g.,~by the ITRS requirements. However, conventional scaling down of MOSFET's channel length is declining as the physical and economic limits of such an approach are coming closer. Novel solutions are increasingly 
being used in MOSFET channel engineering. The growth of a strained Si layer on a relaxed 
Si$_{1-y}$Ge$_{y}$ buffer layer is a typical technique used to improve the electrical performances of 
MOSFET's device, due to the expected enhancements in the carrier
mobility~\cite{ref:21,ref:22}
 of
such a strained layer. High-performance strained Si$_{1-x}$Ge$_{x}$ 
transistors have been recently obtained and this technology could feature in
future chip generation with channel size of 32 nm or less~\cite{ref:21,ref:Morris}.  
When modeling the electrical currents of such devices, it is required to take into account the fundamental transport properties of the charge carriers that are governed by the structure of the electronic energy bands of the strained material.

Material science computational methods for the calculation of the electronic energy band structure 
fall into two general categories. 
The first category includes the ab initio methods, such as Hartree-Fock or Density Functional Theory 
(DFT), which calculate the electronic structure from first principles, i.e.,~without the need for 
empirical parameters. The second category consists of far more computationally efficient semiempirical 
methods such as the Empirical Pseudopotential Method (EPM), the
Tight Binding (TB) method and the \kp\ method. 
   
Over the past decades, the Local Density Approximation (LDA) 
 variant of  DFT~\cite{ref:2} has been established as a very powerful tool for 
studying the elastic properties and the deformation potentials of strained 
semiconductors~$\cite{ref:nielsen,ref:Walle0,ref:Walle,ref:levine}$.
More recently, the GW many-body-correction to the LDA DFT\cite{ref:Hedin} has yielded semiconductors 
band structures that feature band gap values near their experimental values.
 Ab initio methods are self-consistent methods which utilize a variant approach to calculate the ground state
energy of a many-body system and thus require large computer resources. 
They can only be used in particular situations of high symmetries and are not suitable for
 calculating the transport properties of large systems with confined edge states.
   
Unlike ab initio approaches, EPM, TB and the \kp\ method involve 
fitting parameters to reproduce the experimental 
energy band gaps, the dielectric response and the carrier effective masses. 
Over the past three decades, the EPM with spin-orbit (SO) corrections has proven to be extremely successful in calculating 
the electronic band structure of relaxed and strained semiconductors with
indirect gap\cite{ref:4,ref:gell,ref:friedel,ref:17,ref:24}. 
Recent works using the TB method have also given accurate 
results\cite{ref:Ma,ref:6}. 
In the \kp\ method, the energy band structure is obtained by a set of 
parameters which represent the energy gaps at $\Gamma$, the momentum matrix 
elements and the strength of the SO coupling. 
The number of energy bands (or levels) that are effectively calculated is 
related to the precision of the results. 
Six-level \kp\ model\cite{ref:13}, eight-level \kp\ model and fourteen-level
\kp\ model\cite{ref:10} describe well the highest valence bands (VBs) and 
the lowest conduction bands (CBs) of semiconductors near the center $\Gamma$ 
of the Brillouin zone, but fail to describe the CBs of
 semiconductors with indirect gap.  
 Low-order \kp\ Hamiltonians need
a small number of parameters (typically less than 10), while high-order \kp\ 
methods\cite{ref:9}, referred by Pollak et al.\cite{ref:pollak} as 
"full-zone" \kp\ methods, require a large number of unknown parameters. 
While it is a straightforward
 matter to work out the energy-band structure at any $k$-point in the 
Brillouin zone once these parameters
have been chosen, it requires effort and skill to come up with a 
satisfactory set of parameters\cite{ref:pollak}. For this reason, the
full-zone \kp\ method has been used rarely and for a few {\it bulk} 
semiconductors only, including Si, 
Ge~\cite{ref:9,ref:richard,ref:Humphreys} 
and $\alpha$-Sn~\cite{ref:pollak},
 while no extensive works were performed for strained semiconductors.
               
In this paper, we propose to extend to strained Si, Ge and SiGe alloys the 
thirty-level \kp\ model that was firstly introduced by Cardona et 
al.~\cite{ref:9}. Cardona's model is based on the 15 orbital states 
referred to $\Gamma$. For the first time, the well-known Pikus and 
Bir~\cite{ref:bir} correction for strained materials has been combined
 within this thirty-level \kp\ formalism.         
The \kp\ parameter optimization strategy is based on a 
conjugate-gradient procedure that uses ab initio simulations but also  a 
large amount of experimental data that is currently available for Si and Ge. 

In the first part of this paper, a series of ab initio DFT-LDA simulations 
that include the GW correction and relativistic effects in strained 
Si$_{1-x}$Ge$_{x}$/Si$_{1-y}$Ge$_{y}$ systems has been performed 
with a view to complete the experimental data and to establish a reference
set of energy bands.

In the second part of this paper, we have determined the \kp\ model coupling 
parameters and the deformation potentials that fit as close as possible 
the first principle results, matching not only the energy levels and the 
carrier effective masses, but also the general shape of the band structure 
of relaxed and strained crystals. 
A simple interpolation between the \kp\ parameters for Si and those for Ge 
has been proposed in order to model SiGe alloys. 

In the third part of this paper, a set of comparison is given with 
experimental data in relaxed and strained Si, Ge, and Si$_{1-x}$Ge$_{x}$ 
alloys.
We show that the present \kp\ model accurately reproduces the overall band 
structure, as 
well as the band shifts, the carrier effective masses and the Density Of States (DOS) {\it vs.} applied strain. A second set of comparisons with the widely used Chelikowsly and 
Cohen non-local EPM \cite{ref:4} has also shown a good agreement with \kp\ simulations.

\section{\label{first}First principle simulation setup}

\subsection{\label{firstbulk}Bulk material}

A series of first principle calculations has been performed in 
Si and Ge to obtain a reference set of energy bands, which can be used later for the \kp\ model parameters optimization. The first principle data presented in this work have been obtained within the LDA variant of the DFT~\cite{ref:3}. The present DFT-LDA calculation
 relies on the pseudopotential (PP) approximation, by which the core states are effectively eliminated from the calculation.
We have used the well-known published Hartwigsen-Goedecker-Hutter relativistic separable
 dual-space Gaussian PP's~\cite{ref:14} that uses Cerperley and Alder exchange-correlation functional~\cite{ref:ceperley}.
These PPs include relativistic effects and provide an accurate description of the top VB in the near $\Gamma$-region,
 a critical region with respect to the hole transport properties in semiconductors.  
        
The value of the equilibrium lattice parameter has been calculated by minimizing the total energy. Any further LDA calculation has used this theoretical value, instead of the experimental one~\cite{ref:fiorentini}, yielding thus a consistent set of zero-pressure reference data.
In Si, we found a$_0$=5.387 $\AA$, and in Ge a$_0$=5.585 $\AA$.
These values agree within $0.75\%$ and $1.33\%$ with the experimental values of 5.431 $\AA$ and 5.658 $\AA$~\cite{ref:11}, respectively. 
     
It is known that the band gaps calculated with the LDA method are generally 
below their experimental values. However, the 
agreement can be greatly improved by the use of the Hedin's GW 
correction~\cite{ref:Hedin}. 
In practice this correction can be applied as a post-DFT
scheme~\cite{ref:aulbur,ref:GW1,ref:1} in a non self-consistent way. In the following work, 
the G$_0$W$_0$ correction of bulk Si and Ge were 
computed on 19 high-symmetry points in the Brillouin zone and added in a perturbative manner to the LDA band structure. One remark deserves notice: In spite of the G$_0$W$_0$ correction, the theoretical lowest CB typically lie within 0.05-0.2 eV of the range of experimental energies observed~\cite{ref:aulbur}. 
In the present work, the G$_0$W$_0$ theoretical indirect gaps have been found to be 1.076 eV in Si (located at $84\%$ away from $\Gamma$ along the 
$\Gamma$-X direction) and 0.64 eV in Ge, which is underestimated by 8$\%$ and
14$\%$, respectively. Our results compare favorably with other LDA-GW results
found in literature (e.g.,~see the extensive 
comparison between first principle calculations summarized in Aulbur et al. review~\cite{ref:aulbur}).  
Eventhough correctly parametrized, LDA-G$_0$W$_0$ results do not match {\it perfectly} with experimental data, 
and significant theoretical work such as vertex correction, self-consistent-GW and the exact treatment of exchange term are currently in progress to further improve DFT results ~\cite{ref:aulbur}. 
These alternative approaches are beyond the scope of the present work. For the purpose of obtaining reference set of energy bands that can be used in the development 
of an optimized \kp\ model, we used the non self-consistent G$_0$W$_0$ 
approximation to correct the band gap problem and we applied a supplementary 
rigid "squizor" shift of 0.09 eV for Si and 0.104 eV for Ge in order to 
obtain the final reference set of energy bands, the GW energy levels~\cite{ref:Star}. 

Aside from the above studies, the accuracy  of the DFT-LDA calculation critically depends on the manner in which the problem is sampled 
numerically~\cite{ref:wei}. We found out good LDA convergence ($\Delta E_k$~$<<$~0.01~eV) using a basis set of approximatively
 1300-1500 plane waves, which corresponds to a cut-off energy of around 22
 Hartrees.~The Brillouin zone was sampled on a 6*6*6 four-fold shifted Monkhorst
 and Pack grid~\cite{ref:15} (i.e., 864 $k$-points) to obtain the charge density. 
As regard G$_0$W$_0$ correction, a satisfactory trade-off between numerical convergence 
and computation time was achieved with a cut-off energy of 8 Hartrees and using a large 
number of bands ($>$ 100) included in the calculation of the self-energy~\cite{ref:tiago}.

\subsection{\label{firstSiGe}SiGe alloys}
In Si$_x$Ge$_{1-x}$, where $x$ denotes the relative mole fraction of the two materials, both Si and Ge atoms are present
 in the unit cell. For this reason, we used a 32-atoms tetragonal cell to simulate SiGe alloys (a 2-atoms orthorombic unit cell was used for Si and Ge). 
The Si and Ge atoms have been randomly distributed in the super-cell and a structural optimization of the atomic positions in the unit cell has been performed. 
A linear interpolation between the GW correction of Si and Ge was used to correct the band gap. This 
latter approximation is reasonable because the GW correction obtained in Si and Ge are effectively very close \cite{ref:aulbur}. 

The experimental lattice parameter in Si$_x$Ge$_{1-x}$ is well described by the Dismukes's law~\cite{ref:dismukes} according to which 
$a_{exp}=5.431+0.2x+0.027x^2$ is a quadratic function of $x$. 
A similar expression has been obtained from our theoretical results at various $x$-content:
\begin{equation}
a_{theo}=5.387+0.1428x+0.0532x^2.
\label{eq:param}
\end{equation}

\subsection{\label{ab}Strained material}

Epitaxial Si$_{1-x}$Ge$_x$ layers grown on a relaxed Si$_{1-y}$Ge$_y$ buffer is studied 
with a view to fitting the deformation potential parameters of the \kp\ model for strained materials that will be presented hereafter. 
A series of first principle simulations of the electronic band structure has been performed for a large range of biaxial 
strain (up to 4 $\%$) applied perpendicularly to the [001], [111] and [110] directions. 

Using continuum elasticity theory, the strain tensor 
in the layer in the case of [100]-buffer writes,

\begin{equation}
\left\{ 
\begin{array}{ll}
\epsilon _{xx} = & \epsilon _{yy}=\epsilon _{||}\\
\epsilon _{zz} = & \epsilon _{\perp}\\
 \epsilon _{xy} = & \epsilon _{xz}=\epsilon _{yz}=0,
\end{array}
\right.
\label{eq:epsilonxx}
\end{equation}

in the case of [110]-buffer,
  
\begin{equation}
\left\{ 
\begin{array}{ll}
\epsilon _{xx} = &\epsilon _{yy}= \frac{1}{2} \left(\epsilon _{\perp} + \epsilon _{||}\right)\\ 
\epsilon _{zz} = &\epsilon _{||}\\
\epsilon _{xy} =  &\frac{1}{2} \left(\epsilon _{\perp} - \epsilon _{||}\right)\\ 
\epsilon _{xz} =  &\epsilon _{yz}= 0,\\
\end{array}      
\right.   
\label{eq:epsilonxx_1}   
\end{equation}  
              
and in the case of [111]-buffer:
\begin{equation}
\left\{ 
\begin{array}{ll}
\vspace{.15 cm}
\epsilon _{xx} = & \epsilon _{yy}=\epsilon _{zz}=\frac{1}{3} ( \epsilon _{\perp} + 2 \epsilon _{||})\\
 \epsilon _{xy} = & \epsilon _{xz}=\epsilon _{yz}=\frac{1}{3} (\epsilon _{\perp}-\epsilon _{||}),\\
\end{array}      
\right.   
\label{eq:epsilonxx_2}   
\end{equation}
where the longitudinal strain $\epsilon _{||}=a_{||}/a_0-1$ depends on the slight difference between the cubic lattice 
parameter in the buffer ($a_0$) and the longitudinal one in the layer ($a_{||}$). These equations depend on the normal 
strain $\epsilon _{\perp}=-D.\epsilon _{||}$. The Poisson ratios D determine 
the displacements of atomic plans along the normal [001], [111] and [110] directions, in order to minimize the normal stress.  
 
The knowledge of the elastic constants of 
the material is not enough to infer the position of the atoms in the unit cell.
An additional degree of freedom (that occurs notably for shear distortion) must be added to 
the displacement of the atoms~\cite{ref:kleinman}. The internal strain parameters $\xi$, which measure how the distance between the two atoms
in the unit cell changes in response to the symmetry-breaking stress, are known experimentally in Si and Ge uniaxially strained along 
the [111] direction~\cite{ref:noteksi} (and are quoted in Table~\ref{tab:elastic}).

\squeezetable
\begin{table}
\caption{\label{tab:elastic}Elastic coefficients and internal displacement.}
\begin{ruledtabular}
\begin{tabular}{lcccccc}
&& \multicolumn{2}{c}{Si}&& \multicolumn{2}{c}{Ge}\\    
 &&Exp.& LDA\footnote{present work; $^b$elastic constant calculated from the D
 values; $^c$C. S. G. Cousins et al., J. Phys. C {\bf 20}, 29 (1987);$^d$ H. J.
 McSkimin,  J. of Appl. Phys. {\bf 24}, 988 (1953), cited by
 Ref.~\onlinecite{ref:Walle}.}&&Exp.&LDA$^a$\\ 
\hline     
 
 $D_{001}$ (GPa)&&0.776&0.795&&0.7513&0.711\\
 $D_{110}$ (GPa)&&0.515&0.527&&0.4498&0.42\\
 $D_{111}$ (GPa)&&0.444&0.461&&0.3711&0.343\\

 $C_{11}$ (GPa)&&167.5$^d$& 168.3$^b$ && 131.5$^d$&132.8$^b$  \\         
 $C_{12}$ (GPa)&&65$^d$& 66.8$^b$&& 49.4$^d$& 46.8$^b$\\    
 $C_{44}$ (GPa)&&80.1$^d$& 79.9$^b$&&68.4$^d$& 66.57$^b$\\
 $\xi$&&0.54$^c$ &0.536&&0.54$^c$&0.495\\
     
\end{tabular} 
\end{ruledtabular}
\end{table}

The normal stresses and the inner displacement of 
atoms in the cell affect the electronic and structural properties of the strained crystal~\cite{ref:umeno}.
The present calculations, that include a structural optimization of the unit cell, are performed as follows:
  
(i) $a_{||}$ is determined from Eq.~\ref{eq:param}. Eqs.~\ref{eq:epsilonxx},~\ref{eq:epsilonxx_1} and~\ref{eq:epsilonxx_2} are used to calculate the shifted Bravais lattice 
$R'=(1+\epsilon)\cdot~R$.
     
(ii) The internal strain parameter and the Poisson ratios are calculated by minimizing the total energy of the biaxially strained crystal. 
For searching the equilibrium structure under external applied strain $\epsilon_{||}$, the total energy 
is minimized by varying the unit cell length along the strain direction and the atomic position in the unit cell.
  
(iii) The optimized cell is used later in the DFT-LDA band structure calculations. 

(iv) The GW correction of {\it bulk} Si and Ge is used 
to correct the bandgap problem. This choice was motivated by the work of Zhu et al.~\cite{ref:zhu} reporting  
that there is no quantitative difference in Si between the LDA band gap pressure dependencies and the ones from full GW calculation.  
                  
The theoretical coefficients D$_{001}$, D$_{111}$ and D$_{110}$, and the internal strain parameter extracted from the previous structural 
optimizations (step ii) at $\epsilon_{||}~\rightarrow~0$ are reported in Table~\ref{tab:elastic}.
Using continuum elasticity theory, these coefficients  write $D_{001}=2 ({C_{12}}/{C_{11}})$, 
$D_{110}={(C_{11}+3 C_{12}- 2 C_{44})}/{(C_{11}+C_{12}+2 C_{44})}$ and $D_{111}= (2 C_{11}+4 C_{12}-4 C_{44}) / (C_{11}+2 C_{12}+4 C_{44})$, where 
$C_{11}$, $C_{12}$ and $C_{44}$ are the elastic constants. 
Also reported in Table~\ref{tab:elastic} 
are the theoretical elastic constants obtained from D$_{001}$, D$_{111}$ and D$_{110}$ together with the above equations. 
As can be seen, the agreement with experimental values is good. 
 
\begin{figure}
\includegraphics[scale=1.05]{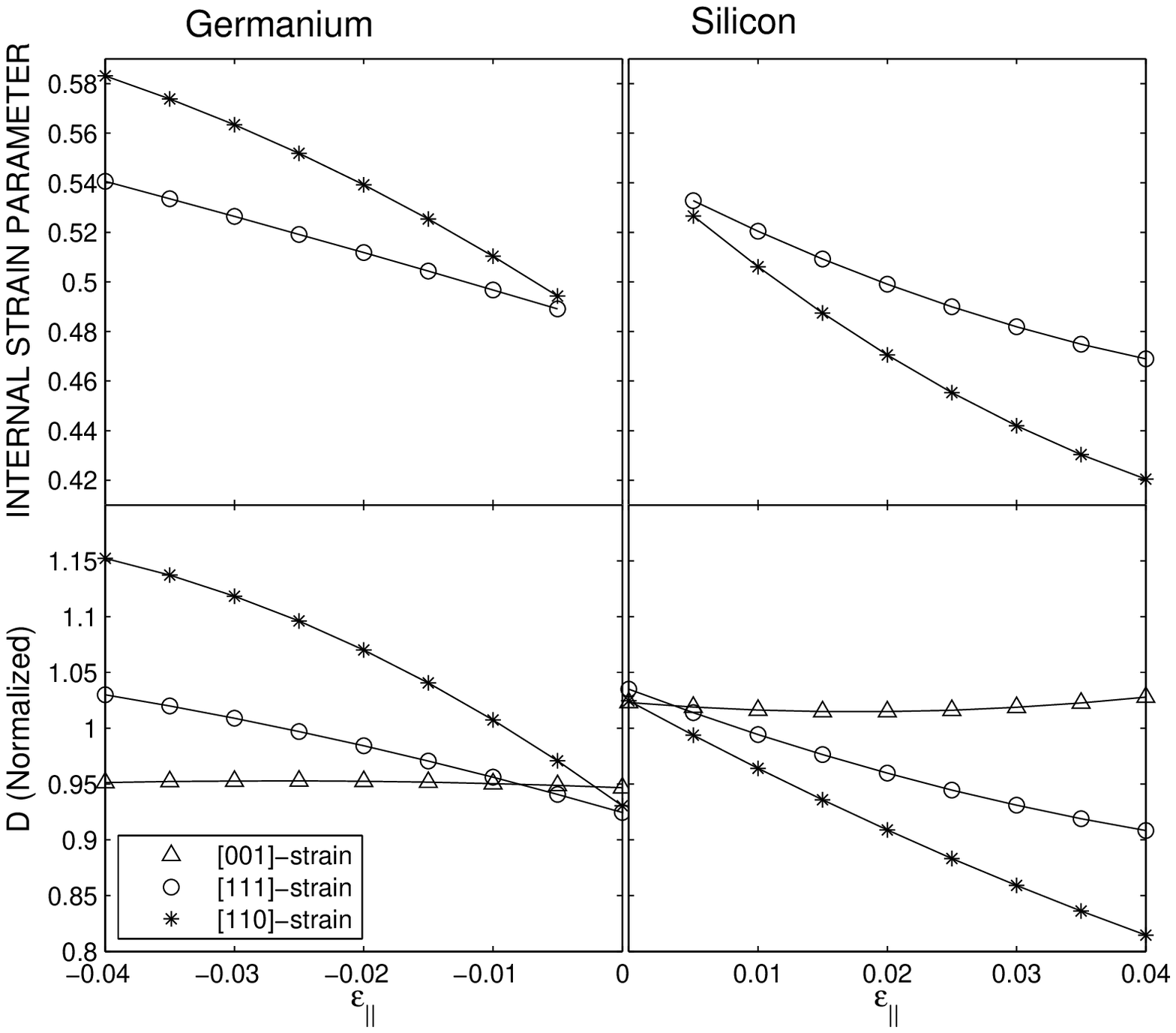}
\caption{\label{fig:atomic_pos} Theoretical internal strain parameter {$\xi$} and theoretical normal strain coefficient D as a function of longitudinal biaxial strain applied perpendicularly to the normal [001], [111] and [110] directions. The coefficients D are normalized to the experimental values listed in Table~\ref{tab:elastic}.}
\end{figure}

In Fig.~\ref{fig:atomic_pos}, the theoretical Poisson ratios D and the internal strain parameters $\xi$ are reported as a function of applied biaxial strain $\epsilon_{||}$. Simulations have been 
performed in tension in Si (right) and in compression in Ge (left) for the three previously mentioned growth cases. 
As can be seen, for the [001]-growth case D does not depend on $\epsilon_{||}$, while for the [111] and [110]-growth cases both D and 
$\xi$ change (up to $\sim$ 15$\%$) with applied strain. The former cases is
consistent with results of Ref.~\onlinecite{ref:Walle0} concerning strained Ge layers grown on [001]-cubic Si. Changes in $\xi$ with applied strain have also been reported in 
 Refs.~\cite{ref:nielsen,ref:levine,ref:Verges} as well as elastic constant strain-dependency~\cite{ref:umeno}. 
As shown later, the energy band shifts notably depend on the deformation applied on the crystal and thus the results shown in Fig.~\ref{fig:atomic_pos} have been carefully accounted for. 
For instance, we found out that the VB energy shifts in Si 
grown on a [111]-Ge buffer are overestimated (by roughly 10$\%$) when both $\xi$ and D are kept constant (this behavior is inversed in Ge grown on [111]-Si buffer).
   
\section{\label{development}Development of An Optimized \kps\ Model}

\subsection{\label{kp}\kp\ parameters for bulk materials}
    
The \kp\ formalism using Zinc-Blende $\Gamma$-centered Bloch function basis $u_{lk}({\bf r})=\sum_{n}C_{n}^{l}u_{n0}({\bf r})$ 
leads to the secular \kp\ equation of the undeformed crystal ~\cite{ref:13}:
\begin{eqnarray}
\sum_{n}&\left\{\left(\frac{\hbar^2k^2}{2m}+E_{n}^{0}-E_{lk}\right)\delta_{n,n'}\right.\nonumber\\
&\left.+\frac{\hbar{\bf k}}{m}\cdot\langle u_{n'0}\left|{\bf p}\right|u_{n0}\rangle \right\} C_{n}^{l}=0,
\label{eq:2}
\end{eqnarray}  
where $E_{n}^{0}$ are the eigenvalues at $\Gamma$. The fifteen $\Gamma$-states from group $O^h$ determined by Cardona et al.~\cite{ref:9} 
(shown in Table~\ref{tab:table3}) is our starting point. The number of independent non-diagonal matrix elements in Eq.~(\ref{eq:2}) can be reduced to ten in Si and Ge using 
group-theory selection rules~\cite{ref:12,ref:9}. 
The SO coupling terms are introduced in the usual way~\cite{ref:13,ref:pollak} leading to a 30$\times$30 \kp\ matrix (See Appendix A).           
One notes incidentally, that since this approach does not use renormalized Luttinger-like parameters, it is different from other lower-order \kp\ models such as 
the fourteen-level \kp\ model of Ref.~\onlinecite{ref:10} and the twenty-level
model of Ref.~\onlinecite{ref:8}.

\begin{table*}
\caption{\label{tab:table3}Eigenvalues and SO splittings of $\Gamma$-centered
states. Symbols of Ref.~\onlinecite{ref:9}. The $\Gamma_{{25'}^l}$ eigenvalue is arbitrarily set to zero. $\Delta$ symbols refer to the state's SO splittings. All energies are expressed in eV.}
\begin{ruledtabular}
\begin{tabular}{lcccccccccc}
States & \multicolumn{3}{c}{Si}&\multicolumn{3}{c}{Ge}&\multicolumn{1}{c}{Si$_{1-x}$Ge$_x$}\\
 at $\Gamma$ & Exp. &EPM \footnote{present calculations; $^b$as presented in
 Ref.~\onlinecite{ref:4}; $^c$as presented in Ref.~\onlinecite{ref:11};
 $^d$Ref.~\onlinecite{ref:ortega}; $^e$as used in the present \kp\ model;
 $^f$Ref.~\onlinecite{ref:lautenschlager}; $^g$Ref.~\onlinecite{ref:vina}; $^h$as presented in
 Ref.~\onlinecite{ref:hybertsen}; $^i$as presented in
 Ref.~\onlinecite{ref:aulbur}.}&
 G$_0$W$_0$$^{ a}$& Exp.&EPM$^{ a}$& G$_0$W$_0$$^a$& \kp$^{ e}$\\
\hline                            
 $\Gamma_{{1}^l}$ &-12.4$\pm$0.6$^b$; -11.2$^i$;&-12.36 & -11.489&
 -12.6$\pm$0.3$^b$;&-12.624 & -12.638& -12.7-0.18$x$ \\
&-12.5$\pm$0.6$^c$;  -11.4$^i$&&& -12.9$\pm$0.2$^h$&&\\
 $\Delta_{{25'}^l}$ & 0.044$^c$&0.044& 0.0499& 0.296$^c$& 0.297& 0.312& 0.044+0.2$x$+0.052$x^{2}$\\     
 $\Gamma_{{15}}$ & 3.4$^c$; 3.35$^f$;&3.406& 3.204&3.006$^i$; 3.206$^i$;&3.279&3.1& 3.335-0.222$x$ \\ 
&3.05$^d$ &&& 3.16$^g$; 3.25$^c$&&&&\\
 $\Delta_{{15}}$ & 0.04$^c$&0.037 & 0.037& 0.200$^c$ &0.205&0.227 & $0.033+0.157x$\\ 
 $\Gamma_{{2'}^l}$ & 4.15$^i$; 4.1$^d$;&4.062 & 3.96& 0.89$^h$; 0.90$^i$&0.861 & 0.715& 4.15-3.26$x$ \\
 &4.185$^c$; 4.21$^i$&&&&&&&&\\
$\Gamma_{{1}^u}$ & &7.561 & 8.308&  &6.072 & 6.82& 8.4-1.6$x$ \\  
 $\Gamma_{{12'}}$ & &9.371 & 8.451&  &8.665 & 9.925& 8.54+1.76$x$ \\
 $\Gamma_{{25'}^u}$ & &12.203 & 11.41 & &11.334 & 11.193 & 11.7-0.34$x$ \\
 $\Delta_{{25'}^u}$ & &0.009 & 0.012&& 0.0558& 0.029& 0.012+0.03$x$\\
 $\Gamma_{{2'}^u}$ & &13.3 & 15.41&& 12.97 & 14.086& 15.8-1.8$x$ \\   
   
\end{tabular}      
\end{ruledtabular}
\end{table*}

As written in appendix A, the thirty-level \kp\ model depends on seven $\Gamma$-centered eigenvalues, four SO coupling coefficients and ten matrix elements. 
In the development of the present optimized \kp\ model, we attempted to fit experimental electronic properties of 
Si, Ge and SiGe alloys as close as possible.      
Currently, there is insufficient detailed experimental information about the band energies to accurately determine all the 
\kp\ parameters, particularly at high energy ($>$ 5 eV). For this reason, we adopted a mixed approach using experimental data when available and ab initio results otherwise. 
These coefficients were fitted using a conjugate-gradient procedure. 
Satisfactory convergence was determined through a least-square error function between \kp\ eigenvalues and GW results evaluated on a dense ($\backsim$ 1000) 
set of $k$-points in the Brillouin zone. A particular care has been paid in the near-$\Gamma$ region and at the CB minima in order to obtain accurate 
description of the curvature masses and Luttinger parameters. 
We also tried as far as possible to reduce the discontinuity in energy at the K, U equivalent points~\cite{ref:pollak}. Due to missing high energy (220) bands
 in the present \kp\, eigenvalues at K and U differ from several meV ($\sim$7 meV). The coupling between (220) states to the other 
thirty-lowest energy states are naturally more pronounced near K, where the lowest (220) eigenvalues are as low as $\backsim$ 4.5 eV. 
For comparison, the lowest (220) eigenvalues are $\gtrsim$ 26 eV at $\Gamma$, $\gtrsim$ 12.5 eV
 at X and  $\gtrsim$ 12 eV at L. After several attempts to remove this discontinuity by changing the \kp\ parameters values only, we found out that no 
satisfactory trade-off between accuracy and continuity in K, U could be
obtained. For this reason, the present \kp\ model has a discontinuity in K,
U~\cite{ref:smoothing} (as in Ref.~\onlinecite{ref:9}).

Following Pollak et al.~\cite{ref:pollak}, the SO strength between $\Gamma_{25^u}$ and 
 $\Gamma_{25^l}$ states were determined by 
imposing the highest VB to be degenerated at X. $\Delta_{25^l}=44$~meV for Si and  $\Delta_{25^l}=290$~meV for Ge are known by experiments, while
 $\Delta_{25^u}$ and $\Delta_{15}$ were obtained from first principle simulations (see Table~\ref{tab:table3}).
             
The $\Gamma$-centered eigenvalues and the coupling parameters obtained from our procedure are listed in Table~\ref{tab:table3} and Table~\ref{tab:table2}. 
One should mentioned that there are two sets of published \kp\ parameters in
Si~\cite{ref:richard,ref:Humphreys} and one set in 
Ge~\cite{ref:richard} based on the early work of Cardona and Pollak~\cite{ref:9}. Although Cardona and Pollak parameters 
set~\cite{ref:9} provides an accurate description of the main CB minima and the top of the VB, it has the limitation of not including the SO coupling (due to 
computational limitations in the mid-sixties). This was recently done in Si and Ge~\cite{ref:richard} together with a new set of parameters. Unfortunately in Ge, the proposed set of parameters 
failed to improve Cardona and Pollak~\cite{ref:9} one in so far as none of the VB reach the edge of the Brillouin zone with zero slope (or average slope) as required by crystal symmetry and the L-valleys minima are very distant from the Brillouin zone edge. This later severe drawback makes this model unappropriate 
for application to transport properties in nanostructure. This was not the case with Cardona and Pollak~\cite{ref:9} parameter set, but also with the present one. In comparison to this former set of parameters, our optimization strategy based on ab initio reference set of energy bands 
brings additional informations (notably at high energies) and slightly improves the accuracy of carrier group velocity at the first and second CB 
minima, but also certain energy gap values in both Si and Ge. Parameters listed in Table~\ref{tab:table3} and Table~\ref{tab:table2} are slightly different from Cardona and Pollak~\cite{ref:9} ones. 
The main differences~\cite{ref:normaliz} come from the fact that the  $\Gamma-$~eigenvalues used in the present \kp\ model differ at high energy (E $>$ 3.5 eV), and that non-local effects have been accounted for in the present model~\cite{ref:non_local}.

\begin{table}    
\caption{\label{tab:table2}Matrix elements of the linear momentum ${\bf p}$
(a.u.) used in the present \kp\ model. Symbols of group O$^h$ are taken from
Ref.~\onlinecite{ref:9}. Other symbols for Si$_{1-x}$Ge$_x$ (0 $<x<$ 1) belong to group T${^d}$.}
\begin{ruledtabular}
\begin{tabular}{lc}
\multicolumn{1}{l}{Matrix elements (a.u.)}& Si$_{1-x}$Ge$_x$\\       
\hline     
\vspace{0.10 cm}
$P$$\equiv\frac{\hbar}{m} $$\langle\Gamma_{{25'}^l}|{\bf p}|\Gamma_{{2'}^l}\rangle$ & 1.22-0.034$x$\\
\vspace{0.10 cm}
$Q$$\equiv\frac{\hbar}{m} $$\langle\Gamma_{{25'}^l}|{\bf p}|\Gamma_{{15}}\rangle$ & 1.0679+0.0068$x$\\
\vspace{0.10 cm}
$R$$\equiv\frac{\hbar}{m} $$\langle\Gamma_{{25'}^l}|{\bf p}|\Gamma_{{12'}}\rangle$ & 0.5427+0.0884$x$\\
\vspace{0.10 cm} 
$P''$$\equiv\frac{\hbar}{m} $$\langle\Gamma_{{25'}^l}|{\bf p}|\Gamma_{{2'}^u}\rangle$ & 0.156-0.0081$x$\\
\vspace{0.10 cm}
$P'$$\equiv\frac{\hbar}{m} $$\langle\Gamma_{{25'}^u}|{\bf p}|\Gamma_{{2'}^l}\rangle$ & -0.008+0.078$x$-0.05$x^2$\\
\vspace{0.10 cm}
$Q'$$\equiv\frac{\hbar}{m} $$\langle\Gamma_{{25'}^u}|{\bf p}|\Gamma_{{15}}\rangle$ & -0.6555-0.1052$x$\\
\vspace{0.10 cm}
$R'$$\equiv\frac{\hbar}{m} $$\langle\Gamma_{{25'}^u}|{\bf p}|\Gamma_{{12'}}\rangle$ & 0.8342-0.0126$x$\\
\vspace{0.10 cm}
$P'''$$\equiv\frac{\hbar}{m} $$\langle\Gamma_{{25'}^u}|{\bf p}|\Gamma_{{2'}^u}\rangle$ & 1.425-0.0263$x$\\
\vspace{0.10 cm}
$T$$\equiv\frac{\hbar}{m} $$\langle\Gamma_{{1}^u}|{\bf p}|\Gamma_{{15}}\rangle$ & 1.166-0.0247$x$-0.04$x^2$\\
\vspace{0.10 cm}
$T'$$\equiv\frac{\hbar}{m} $$\langle\Gamma_{{1}^l}|{\bf p}|\Gamma_{{15}}\rangle$ & 0.29+0.08$x$\\    
\vspace{0.10 cm}
$S$$\equiv\frac{\hbar}{m} $$\langle\Gamma_{{15}}|{\bf p}|\Gamma_{{2'}^l}\rangle$ & -i0.1$x(1-x)$\\    
\vspace{0.10 cm}
$S'$$\equiv\frac{\hbar}{m} $$\langle\Gamma_{{15}}|{\bf p}|\Gamma_{{2'}^u}\rangle$ & i0.3$x(1-x)$\\     
                         
\hline   
\hline   
\multicolumn{1}{l}{SO coupling strength (eV)}& Si$_{1-x}$Ge$_x$\\   
\hline  
$\Delta_{\Gamma_{25'^{l}},\Gamma_{25'^{u}}} $ & 0.022+0.198$x$\\
$\Delta_{\Gamma_{15},\Gamma_{25'^{l}}} $ & 0.04$x$-0.04$x^2$\\
                                 
 \end{tabular}
\end{ruledtabular}
\end{table}

\begin{figure}
\includegraphics[scale=0.95]{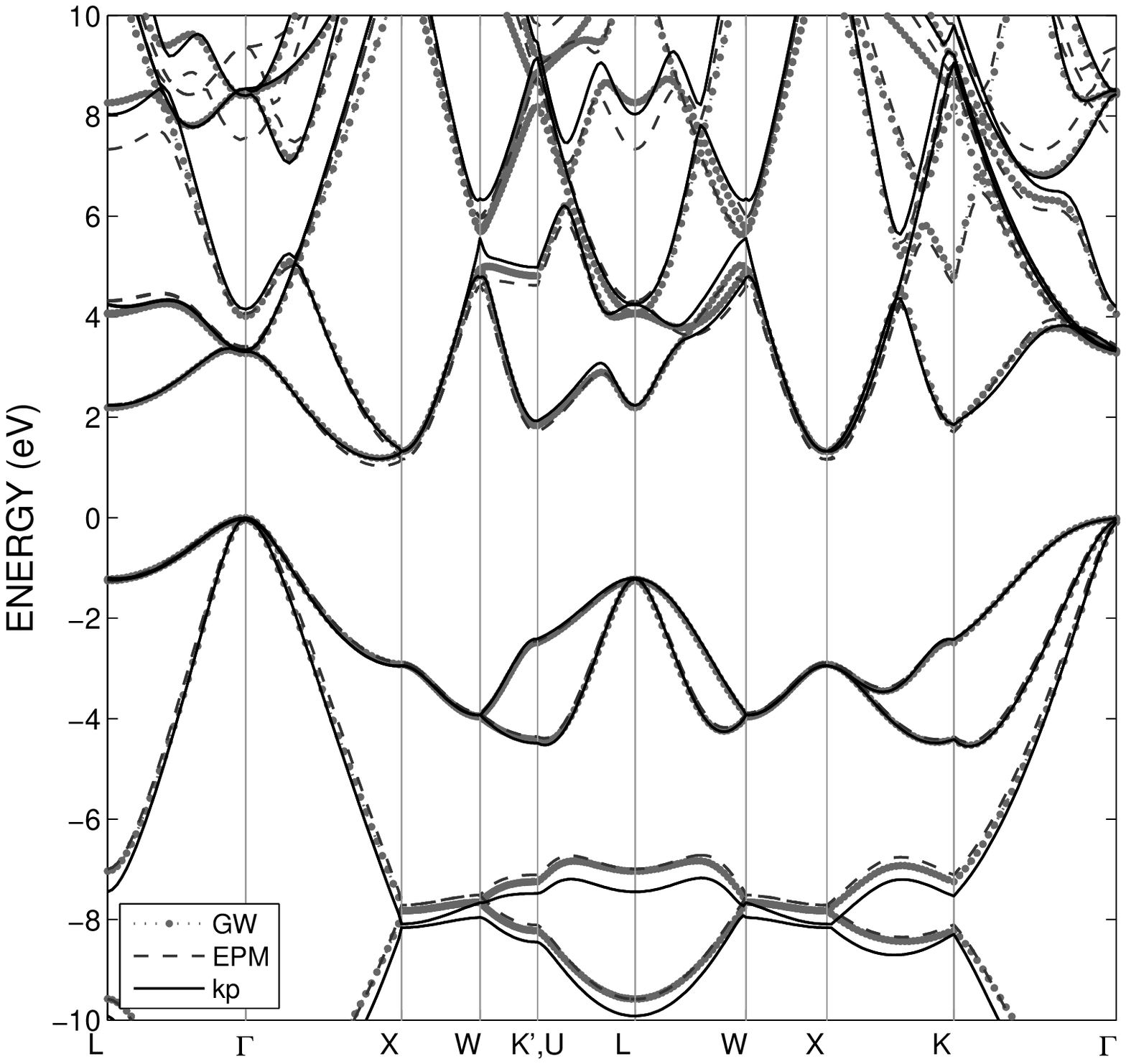}
\caption{\label{fig:BS_si} Bulk Si electronic band structure obtained using thirty-level \kp\ model, EPM and GW calculation.}
\end{figure}

\begin{figure}
\includegraphics[scale=.95]{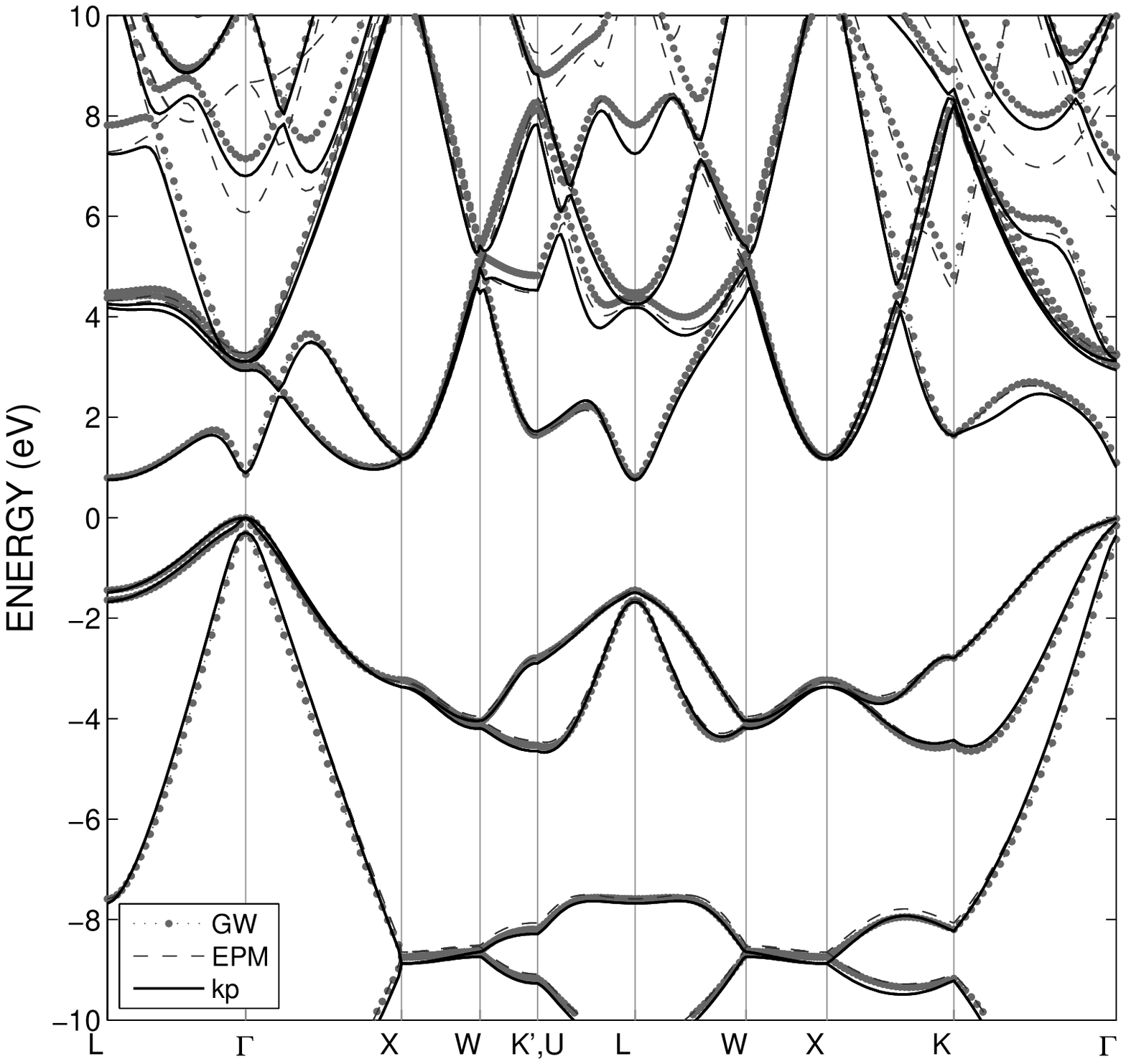}
\caption{\label{fig:BS_ge} Bulk Ge electronic band structure obtained using thirty-level \kp\ model, EPM and GW calculation.}
\end{figure}

The Si and Ge BS obtained using the present \kp\ model are compared to GW first principle simulations in Fig.~\ref{fig:BS_si} and Fig.~\ref{fig:BS_ge}.                
The overall quality of the fit is excellent: 
The difference in band energies between our semiempirical values and those used for
the fit was typically less than 0.01 eV for the principal band gaps, and under 0.3 eV 
at other high symmetry points.
   

Further comparisons are shown in Fig.~\ref{fig:BS_si} and Fig.~\ref{fig:BS_ge} with the widely used Chelikowsky and Cohen non-local EPM~\cite{ref:4}. 
As can be seen, the straightforward application of the EPM yields a BS that is in very good agreement (up to more than 6eV) with the results of the much more complex GW ab initio calculation. 
This excellent agreement means that both methods are consistent and leads further support to the 
GW reference set of energy bands used for the \kp\ parameters optimization.

\subsection{\label{sige} SiGe alloys parameters}  
\begin{figure}
\includegraphics[scale=1.3]{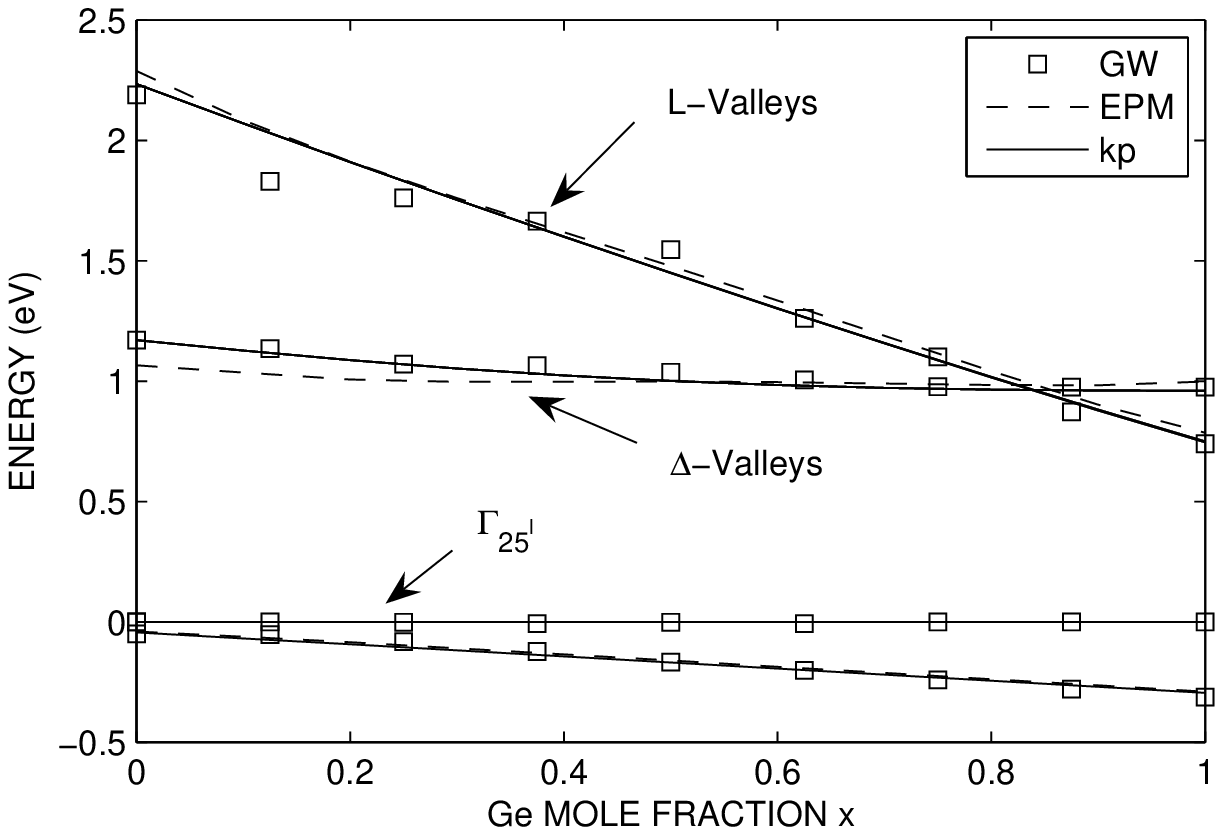}
\caption{\label{fig:gap_sige} Relaxed Si$_{1-x}$Ge$_{x}$ band gaps and SO splittings obtained using thirty-level \kp\ model, EPM and GW calculation.}
\end{figure}

\begin{figure}
\includegraphics[scale=0.95]{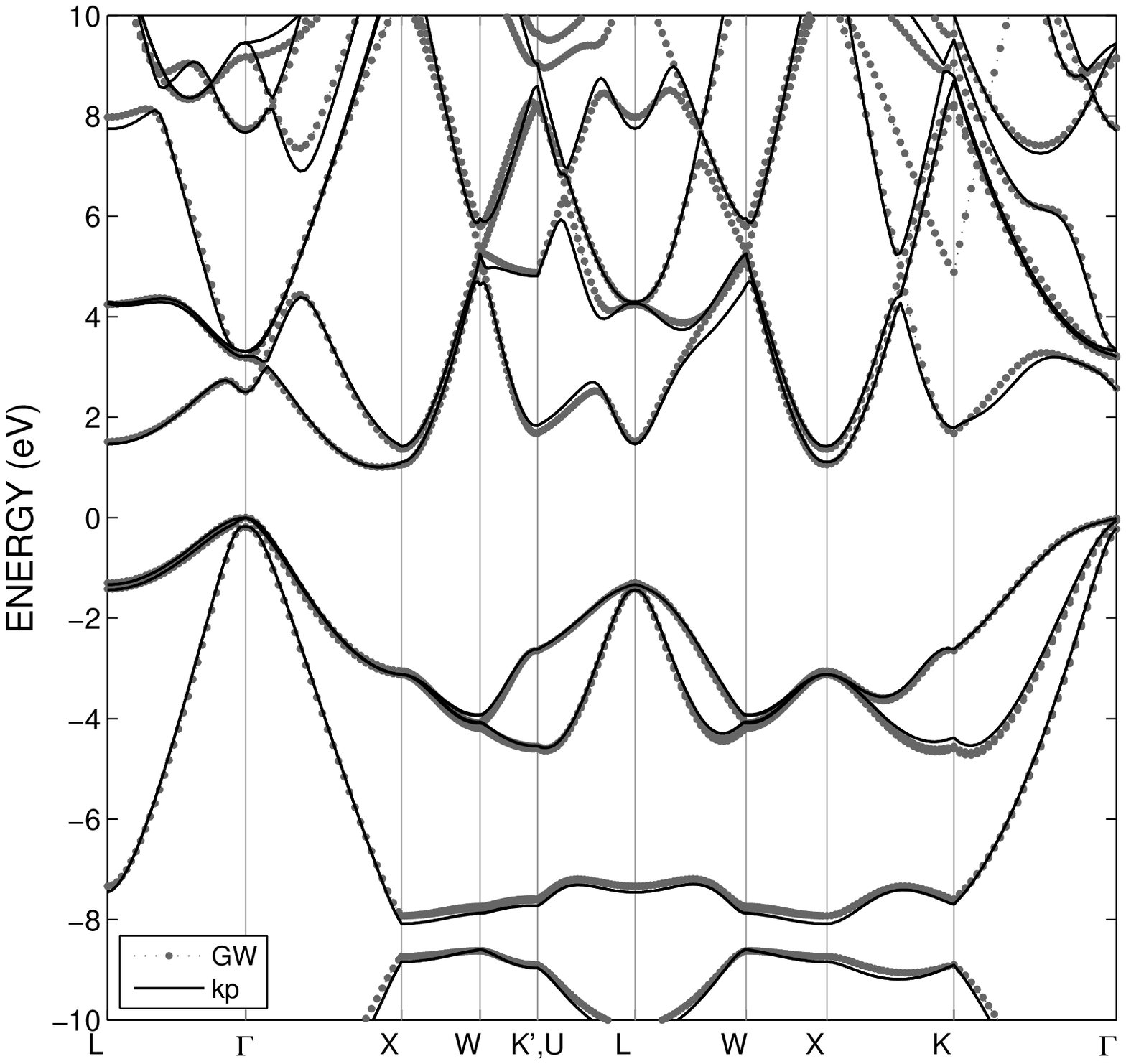}
\caption{\label{fig:BS_sige} Bulk Si$_{0.5}$Ge$_{0.5}$ electronic band structure obtained using thirty-level \kp\ model and GW calculation.}
\end{figure}

The virtual crystal approximation was used to extend the \kp\ results to SiGe alloys. A quadratic interpolation between Si and Ge parameters is proposed in Tables~\ref{tab:table3} and~\ref{tab:table2}.
Due to centro-symmetry breaking SiGe alloys do not belong to group O$^h$. 
A supplementary SO coupling term~\cite{ref:schmid} and two purely imaginary coupling terms from group T$^d$~\cite{ref:kittel} (not shown in appendix A) have been introduced in Table~\ref{tab:table2}.
These interpolation coefficients were determined in order to account for the GW Si$_{1-x}$Ge$_{x}$ band structures for various $x$-content values. Fig.~\ref{fig:gap_sige} 
shows the Si$_{1-x}$Ge$_x$ SO splittings and band gaps as a function of $x$ calculated using the present \kp\ model, but also with EPM and GW model. 
The present \kp\ model predicts a crossing between $\Delta$-valley and L-valley minima at x=0.84 which is consistent with the experimental data~\cite{ref:21}.       
With these new parameters and interpolation coefficients, one obtains a good agreement between the \kp\ model and GW results as it is testified by the Si$_{0.5}$Ge$_{0.5}$ band structure shown in Fig.~\ref{fig:BS_sige}.
       
\subsection{Strained materials: deformation potentials}
       
The analysis of the structure of strained semiconductors using \kp\ Hamiltonian has been initially proposed by
Pikus and Bir~\cite{ref:bir}. Authors derived 
the first-order  
\kp\ perturbation terms arising from straining the semiconductor in question.
This general expression has been widely used in six-level \kp\ Hamiltonian to analyze the effect of deformation on the hole energy spectrum. 
For the first time, we applied Pikus and Bir formalism in the case of the
thirty-level \kp\ Hamiltonian. The perturbation 
term to be added to Eq. \ref{eq:2} writes~\cite{ref:bir}:  
\begin{eqnarray}
W_{n',n} = -\frac{\hbar}{m} &\sum _{i,j} \epsilon _{ij} k_i . <u_{n'0}|p_j|u_{n0}> \nonumber\\
&+\epsilon _{ij}<u_{n'0}|\Xi _{ij}|u_{n0}>,
\label{eq:strcor}
\end{eqnarray}  
where $i,j$ stand for ${x,y,z}$. The first term of Eq. \ref{eq:strcor}, accounts for the interaction between the strain
and the momentum of the carriers. 
The deformation potential operator~{$\Xi_{ij}=-\frac{p_i p_j}{m}+V_{ij}(\bf
r)$} describes the change in the potential and the kinetic energy of carriers due
to the strain itself. The unknown non-vanishing deformation potentials at $\Gamma$ $<u_{n'0}|\Xi_{ij}|u_{n0}>$, listed in Table~\ref{tab:coupkp}, were determined
from group theory selection rules taking into account the $p_i p_j$ contribution only. 
This choice is motivated by the fact that in the deformed-ion 
approximation the V$_{ij}$ term with rhombic or tetragonal symmetries vanishes \cite{ref:bir}.

\begin{table*}
\caption{\label{tab:coupkp}Strain perturbation matrix coefficients expressed in eV.}
\begin{ruledtabular}
\begin{tabular}{cccccccc}
symbols & Si$_{1-x}$Ge${_x}$&symbols &Si$_{1-x}$Ge${_x}$&symbols &Si$_{1-x}$Ge${_x}$&symbols &Si$_{1-x}$Ge${_x}$\\
\hline
 $l_{\Gamma_{25^l}}$ &-2.7-1.1$x$ &$a_{12}$ & 7.7-0.885$x$ &
 $l_{\Gamma_{25^l},\Gamma_{25^u}}$& -19.8-4.339$x$&$a_{\Gamma_{2^l},\Gamma_{2^u}}$& 0.3-1.511$x$\\
 $m_{\Gamma_{25^l}}$ & 4.2+0.7$x$ &$b_{12}$ & 5.47+1.328$x$ &
 $m_{\Gamma_{25^l},\Gamma_{25^u}}$& 3.9-4.024$x$&$a_{\Gamma_{1^l},\Gamma_{1^u}}$& -2-3.927$x$\\
 $n_{\Gamma_{25^l}}$ & -7.379-2.148$x$& $c_{12}$ & 7.3+0.445$x$&
 $n_{\Gamma_{25^l},\Gamma_{25^u}}$& -0.112$x$&$ g_{\Gamma_{12},\Gamma_{2^u}}$& -10.5+5.5$x$\\
 $l'_{\Gamma_{15}}$ & 3.4+2.626$x$& $d_{12}$ & 3.65+1.208$x$&
 $f_{\Gamma_{1^u},\Gamma_{25^u}}$& 6+5.22$x$&$g_{\Gamma_{12},\Gamma_{2^l}}$& -4.5-0.854$x$\\
 $m'_{\Gamma_{15}}$ & -0.5+1.262$x$&$a_{\Gamma_{2^l}}$& -9+1.819$x$ &
 $f_{\Gamma_{1^l},\Gamma_{25^l}}$& -5-2.666$x$&\\   
 $n'_{\Gamma_{15}}$ & -10.392+0.258$x$&$a_{\Gamma_{2^u}}$& 5-0.51$x$&
 $f_{\Gamma_{1^u},\Gamma_{25^l}}$&-10-2.21$x$& \\  
 $l''_{\Gamma_{25^u}}$ & -19-1.692$x$& $a_{\Gamma_{1^l}}$& 10+4.171$x$&
 $f_{\Gamma_{15},\Gamma_{2^l}}$& -19-3.242$x$& \\
 $m''_{\Gamma_{25^u}}$ & 8+1.119$x$&$a_{\Gamma_{1^u}}$& 0.5-0.992$x$&
 $f_{\Gamma_{15},\Gamma_{2^u}}$& -2+21.925$x$& \\  
 $n''_{\Gamma_{25^u}}$ & -1.732+2.213$x$& && \\
\end{tabular}
\end{ruledtabular}
\end{table*}

The perturbation matrix $W$ to be 
added to the thirty \kp\ matrix is shown in appendix B.          
Accurate knowledge of the deformation potentials at $\Gamma$ for {\it all} the thirty 
lowest energy bands is required for the construction of the \kp\ model in strained material. However, only deformation potential for the $\Gamma_{25^l}$ states are known experimentally \cite{ref:Walle0}. 
The deformation potentials, needed in Eq. \ref{eq:strcor} and listed in Table~\ref{tab:potdef} 
have been fitted using a procedure similar to the one used in bulk materials. 
We have used a least square optimization procedure on GW energy levels calculated for various strain tensors, including shear distortions. 
Special attention has been given to the respect of time-reversal symmetry degeneracy at Brillouin zone edge.
In addition to space-group symmetry operations, the Hamiltonian of an isolated centrosymetric crystal exhibits time-reversal symmetry. 
Therefore, additional degeneracy among eigenvalues may be determined using the Kramer theorem and the Wigner rule. Using that
rule, Ma et al.~\cite{ref:Ma} have obtained the additional informations about degeneracy in Si and Ge for the [001], [111], and
[110] growth cases that have been accounted for in the present \kp\ model.

\begin{table} 
\caption{\label{tab:potdef}Deformation Potentials.}
\begin{ruledtabular}
\begin{tabular}{lcccc}
Si &Exp.\footnote{Cited by Ref.~\onlinecite{ref:Walle0}; $^b$ Chelikowsky and
Cohen-based EPM~\cite{ref:4,ref:ourepm}; $^c$present work; $^d$present model.}&EPM$^{ b}$&LDA$^{ c}$&\kp$^{ d}$\\
 $b_v$ &-2.10$\pm$0.10&-2.12&-2.27&-2.27\\	    
 $d_v$ &-4.85$\pm$0.15&-4.56&-4.36&-4.36\\	    
 $\left(\Xi_d^{\Delta}+\frac{1}{3} \Xi^{\Delta}_u-a_v\right)$ &1.50$\pm$0.30& 2.24& 1.67&1.94\\
 $\left(\Xi_d^{L}+\frac{1}{3} \Xi_u^{L}-a_v\right)$ &&-1.4&-3.14&-3.03\\
 $\Xi^{\Delta}_u$ &8.6$\pm$0.4& 9& 8.79&9.01\\
\vspace{.15 cm}   
 $\Xi^{L}_u$ &&15.9& 13.85&15.1\\  
\hline
\vspace{.19 cm}            
Ge  &Exp.$^{ a}$ &EPM$^{ b}$&LDA$^{ c}$&\kp$^{ d}$\\
 $b_v$ &-2.86$\pm$0.15& -2.81& -2.9&-2.8\\        
 $d_v$ &-5.28$\pm$0.50&  -5.31&  -6&-5.5\\          
 $\left(\Xi_d^{\Delta}+\frac{1}{3} \Xi_u^{\Delta}-a_v\right)$ & & 3.12& 1.43&1.83\\
 $\left(\Xi_d^{L}+\frac{1}{3} \Xi_u^{L}-a_v\right)$ &-2.0$\pm$0.5& -2.26& -2.86&-1.97\\
 $\Xi^{\Delta}_u$ &&9.91& 10& 10\\ 
 $\Xi^{L}_u$ &16.2$\pm$0.4&  16.3& 17& 16.3\\    
\end{tabular}  
\end{ruledtabular}  
\end{table}

\begin{figure}
\includegraphics[scale=0.95]{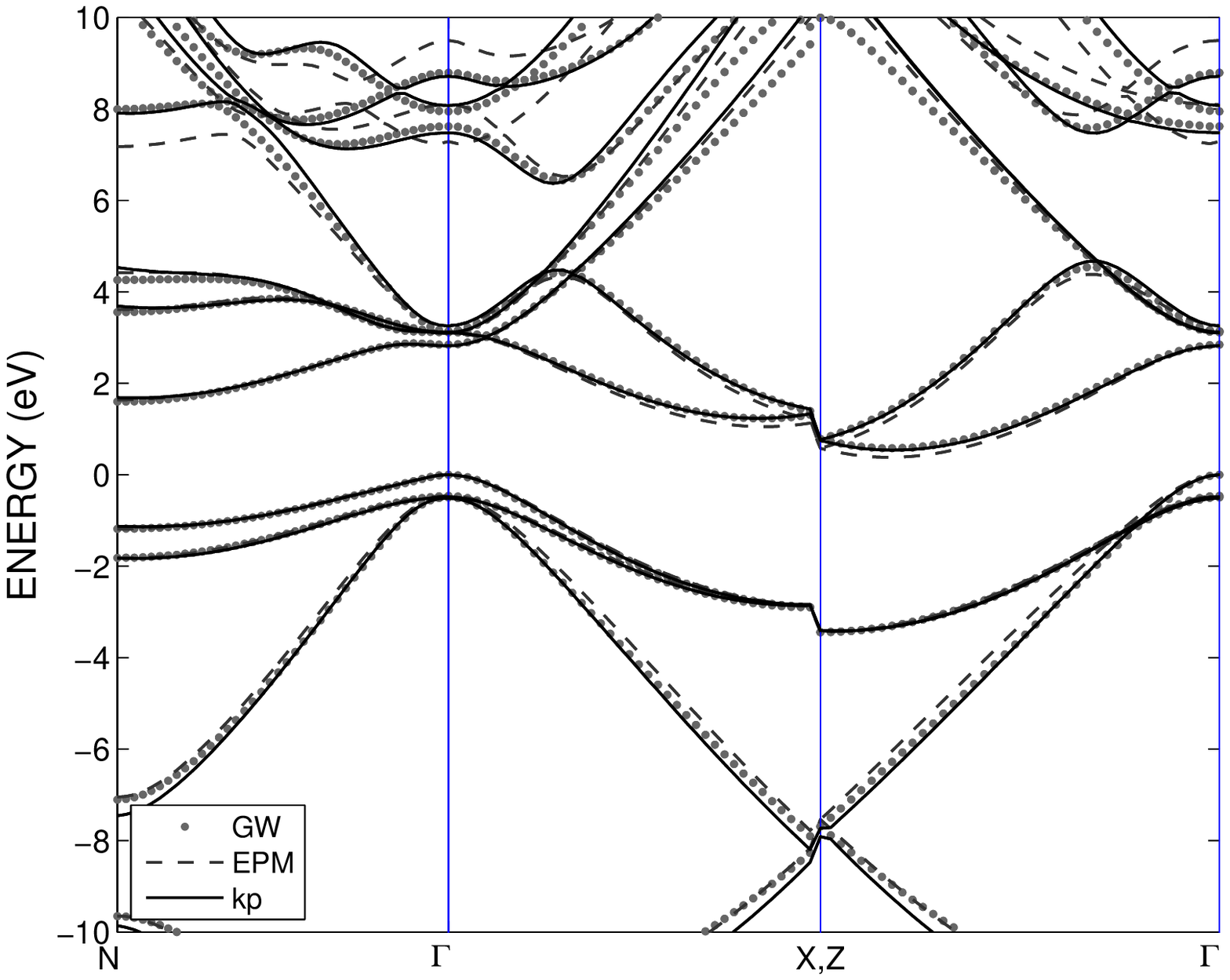}
\caption{\label{fig:BS_si_str} Electronic band structure of strained Si layer grown on [001]-oriented cubic Ge buffer. Simulations with thirty-level \kp\ model (solid lines), EPM (dashed lines) and GW (dotted lines).}
\end{figure}

\begin{figure}
\includegraphics[scale=0.95]{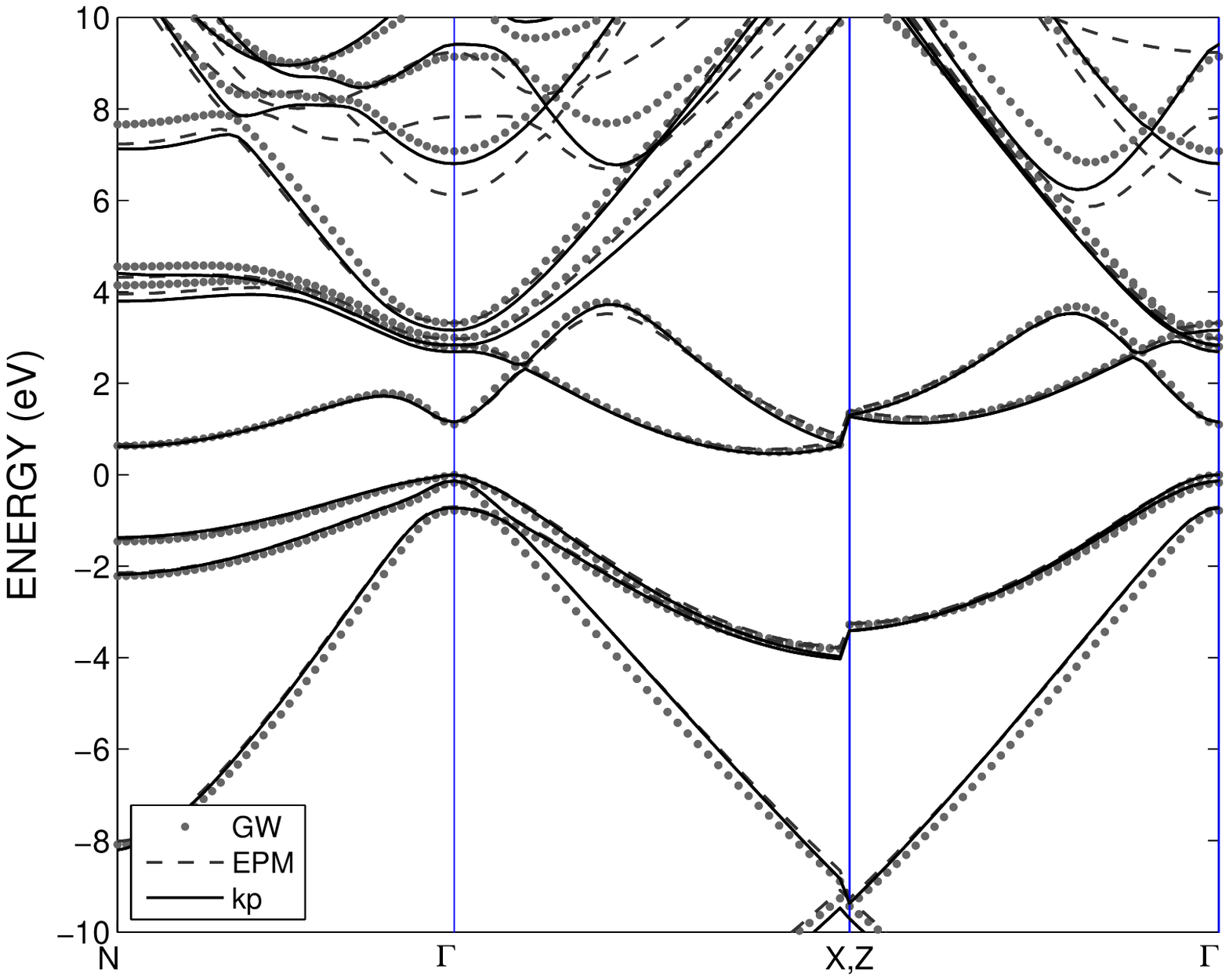}
\caption{\label{fig:BS_ge_str} Electronic band structure of strained Ge layer grown on [001]-oriented cubic Si buffer. Simulations with thirty-level \kp\ model (solid lines), EPM (dashed lines) and GW (dotted lines).}
\end{figure}

\begin{figure}
\includegraphics[scale=0.95]{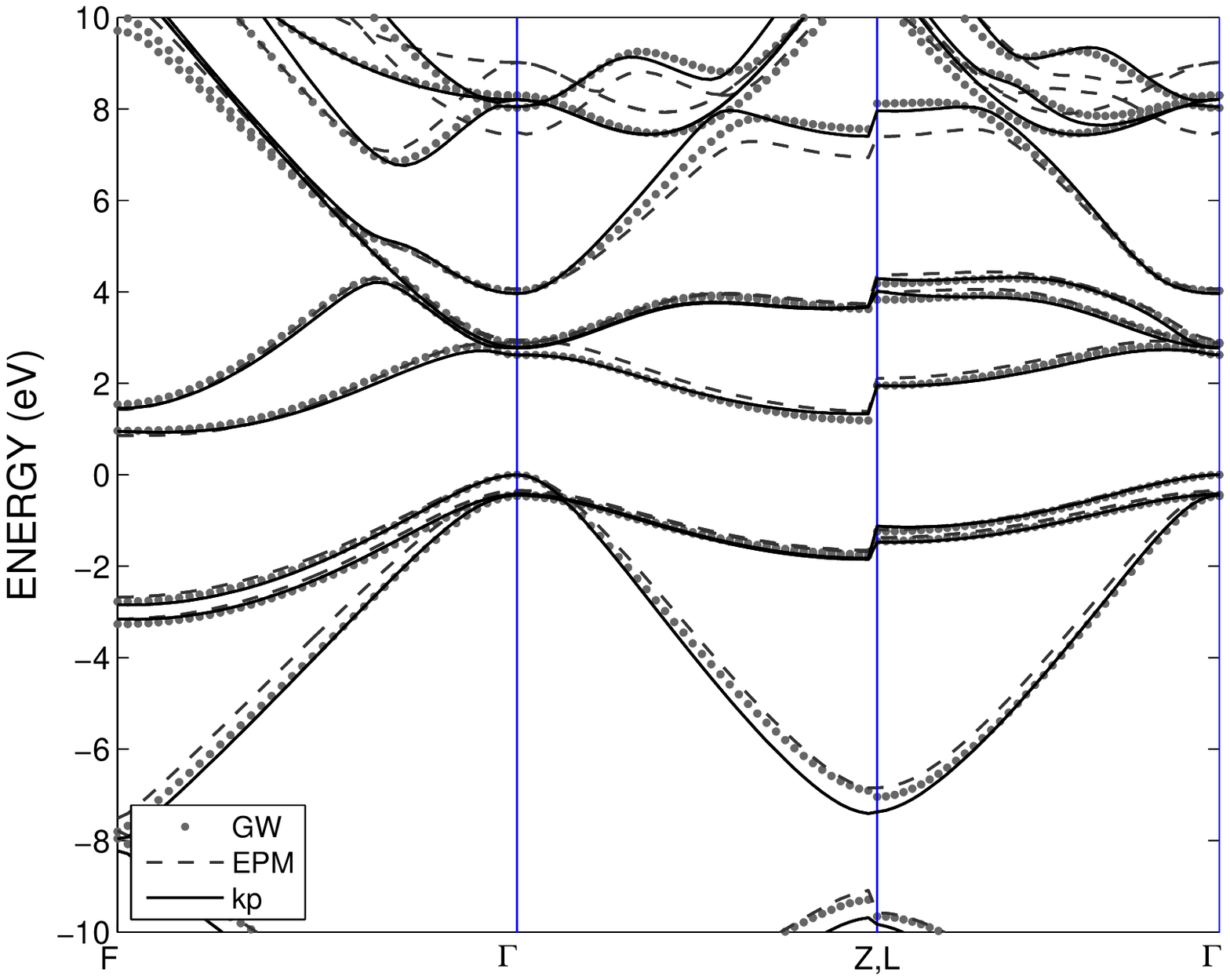}
\caption{\label{fig:BS_si_sh_str} Electronic band structure of strained Si layer grown on [111]-oriented cubic Ge buffer. Simulations with thirty-level \kp\ model (solid lines), EPM (dashed lines) and GW (dotted lines).}
\end{figure}

\begin{figure}
\includegraphics[scale=0.95]{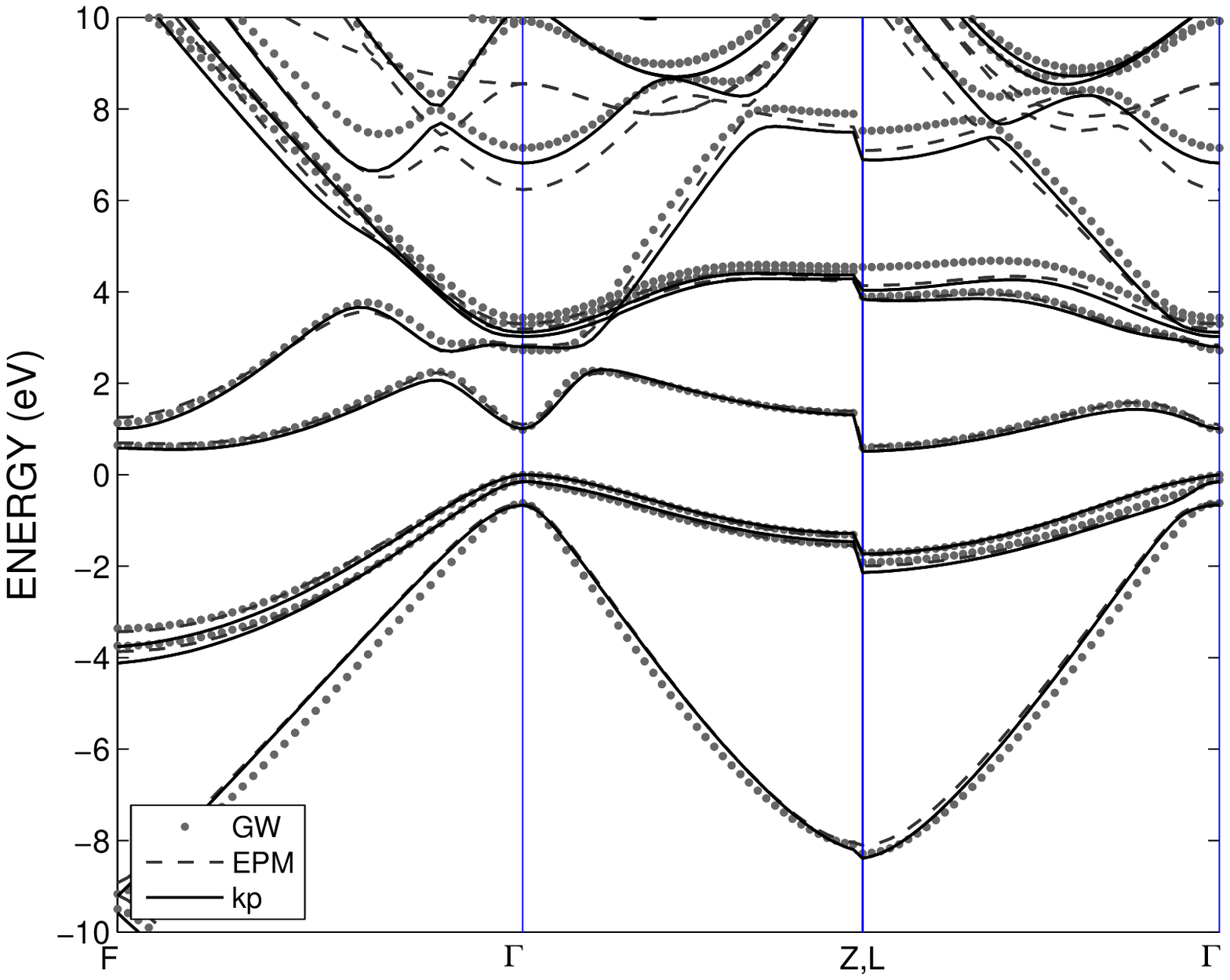}
\caption{\label{fig:BS_ge_sh_str} Electronic band structure of strained Ge layer grown on [111]-oriented cubic Si buffer. Simulations with thirty-level \kp\ model (solid lines), EPM (dashed lines) and GW (dotted lines).}
\end{figure}

Typical results of our fit procedure are presented in Fig.~\ref{fig:BS_si_str} (and Fig.~\ref{fig:BS_ge_str}) for biaxially strained Si (Ge) layers on [001]-oriented cubic Ge buffer (and Si buffer, respectively).
Further results are shown in Fig.~\ref{fig:BS_si_sh_str} (and Fig.~\ref{fig:BS_ge_sh_str}) for strained Si (Ge) layers on [111]-oriented cubic Ge buffer (and Si buffer). 
 The band structures calculated along various directions in reciprocal space using the present \kp\ model are compared to  EPM~\cite{ref:ourepm} and  GW results. 

Under uniaxial strain along [100], the minima of the L-valleys remain equivalent.
 However, due to crystal symmetry lowering by biaxial-strain, the VB degenerate levels are 
split and equi-energy lowest CB $\Delta$-valleys are split into four $\Delta_4$ and two $\Delta_2$-valleys.               
On the contrary, uniaxial strain along [111] leaves the $\Delta$-valleys equivalent whereas valleys at second 
minima are split (two Z-valleys along [111] and six L-valleys along 
[$\overline 1$11], [1$\overline 1$1]  and [11$\overline 1$]).

Another important effect of the strain is observable in the near-$\Gamma$ region: In the case of the strained Si (tensile-biaxial-strain),  
the heavy-hole band "crosses" the SO band, while in the case of strained Ge it is the contrary (compressive-biaxial strain); The SO-hole band "crosses" the heavy-hole one.

\section{\label{experimental1}Results and comparison to experimental data}
 \subsection{\label{experimental} Energy band gaps}

We address now the question, how well do the energy levels ({\it vide post} for the effective masses) calculated using \kp\ agree with the experiments and theoretical results. 
In Tables~\ref{tab:table3} and~\ref{tab:highsym} comparison is given for the cases where experimental data seem to be well
 established.  To this end, we have used the large set of experimental values summarized by Chelikowsky and Cohen~\cite{ref:4}, Landolt-Bornstein~\cite{ref:11}, Hybertsen and Louie~\cite{ref:hybertsen}, Aulbur et al.~\cite{ref:aulbur}, 
but also the most recent optical measurements in Si~\cite{ref:lautenschlager} and Ge~\cite{ref:vina}.
One notes that this assignment is in some cases somewhat tentative. The interpretation of the experimental peaks and critical points 
is hindered when energetically close transitions take place, e.g., due to SO splitted bands. Moreover, the critical points might originate from transitions close to, but not necessarily exactly at high symmetry points in the Brillouin zone.
Aside from recent inverse-photoemission and photoemission experiments which have addressed certain high energy bands in Si and Ge~\cite{ref:ortega}, 
the complex shape of the different energy bands cannot be easily measured directly on its whole structure.

\begin{table*}    
\caption{\label{tab:highsym} Eigenvalues and energy gaps at high-symmetry points in the Brillouin zone calculated with EPM, \kp\ and GW methods (see text for details). The state's SO splittings are shown in parenthesis. Averaged values over transitions between SO splitted bands are noted with a bar. All energies are in eV.}
\begin{ruledtabular}    
\begin{tabular}{lcccccccccc}    
States \footnote{symbols of Ref.~\onlinecite{ref:kittel}; $^b$as presented in
Ref.~\onlinecite{ref:4}; $^c$as presented in Ref.~\onlinecite{ref:ortega};
$^d$as presented in Ref.~\onlinecite{ref:11};
$^e$Ref.~\onlinecite{ref:lautenschlager}; $^f$Ref.~\onlinecite{ref:vina}; $^g$as
presented in Ref.~\onlinecite{ref:hybertsen}.} &
\multicolumn{4}{c}{Si}&\multicolumn{4}{c}{Ge}\\    
&Exp.&EPM & G$_0$W$_0$ & \kp&Exp.&EPM& G$_0$W$_0$ &  \kp\\
\hline      
 $L_1$ &-6.8$\pm$0.2$^b$; -6.4$^i;$& -6.991 & -7.019 & -7.448 &-7.7$\pm$0.2$^b$& -7.588 & -7.801 & -7.678\\
 & -6.7$\pm$0.2$^d$&&&&&&&&\\
 $L_3'$&-1.5$^c$; -1.2$\pm$0.2$^b$& -1.228 & -1.216 & -1.198&-1.4$\pm$0.2$^b$&-1.433 & -1.459& -1.490\\  
 &&(0.034)&(0.033)&(0.026)&&(0.187)&(0.197)&(0.188)\\
 $L_1$ &2.06$^i$; 2.1$^g$;& 2.247 & 2.095 & 2.234 &0.744$^d$& 0.776& 0.64&0.747\\
  & 2.4$\pm$0.15$^g$ &&&&&&&&\\
$L_3$ &3.9$^i$; 4.15$\pm$0.1$^g$& 4.324 & 3.962 & 4.245 &4.2$\pm$0.1$^g$; 4.4$^c$;&4.319 &4.227  & 4.250 \\
 &&&&& 4.3$\pm$0.2$^d$ &&&\\
 &&(0.016)&(0.015)&(0.007)&&(0.087)&(0.103)&(0.077)\\
 $L_2'$&&7.334 & 8.161 & 8.031&7.8$^c$; 7.8$\pm$0.1$^g$;& 7.285 & 7.495& 7.242\\
 &&&&&7.9$^i$&&&\\
 $X_1$ &&-7.711 &-7.823 &-8.087 &-9.3$\pm$0.2$^g$&-8.646 & -8.995& -8.875\\
 $X_4$ &-2.5$\pm$0.3$^b$; -2.9$^g$;&-2.889 &-2.92 &-2.95 &-3.66$^c$; -3.15$\pm$0.2$^d$;&-3.267 & -3.28 & -3.375\\
 & -3.3$\pm$0.2$^g$&&&& -3.5$\pm$0.2$^g$&&&\\
 $X_1$ &1.13$^i$; 1.25$^c$; 1.3$^g$& 1.163 & 1.221 & 1.321 &1.3$\pm$0.2$^d$& 1.254 & 1.045 & 1.169 \\
 $W_1$ & -8.1$\pm$0.3$^b$ &-7.512   & -7.653  & -7.662 &-8.7$\pm$0.3$^b$&-8.512 & -8.88 & -8.638 \\
 &&(0.006)&(0.005)&(0.295)&&(0.029)&(0.025)&(0.103)\\
 $W_2$ & -3.9$\pm$0.2$^b$ & -3.886 &-3.95  &-3.922 & -3.9$\pm$0.2$^b$ &-3.956  &-4.151  & -4.038 \\
 &&(0.014)&(0.013)&(0.008)&&(0.042)&(0.075)&(0.154)\\
 $\Sigma_{1}^{min}$ &-4.7$\pm$0.2$^b$&-4.466&-4.527&-4.553&-4.5$\pm$0.2$^b$&-4.548&-4.748&-4.555& \\
 $E_g\left(\Delta\right)$ &1.17$^d$&1.031 &1.076 & 1.17 &&1.04 &0.855 &0.961\\
 $\overline{E_1}(L)$ & 3.45$^g$; 3.46$^e$&3.492&3.311 &3.432 &2.05$^d$; 2.22$^f$&2.302  &2.099&2.239\\
 $\overline{E'_1}(L)$ &5.38$^e$; 5.50$^g$&5.577 &5.178 &5.443  &5.65$^d$&5.889& 5.686&5.750 \\
 $E_2(X)$ &4.32$^e$&4.052 &4.141 &4.271  &4.45$^f$&4.521&4.325 &4.544 \\
              
\end{tabular}
\end{ruledtabular}
\end{table*}

It is clear from Tables~\ref{tab:table3} and~\ref{tab:highsym} that the overall agreement between theoretical and experimental band gaps is good. 
In particular, the present \kp\ model predicts indirect gaps of 1.17 eV in Si and 0.747 eV in Ge.

\subsection{\label{curvature} Curvature masses, Luttinger parameters and DOS}

The $\Delta$-electron and L-electron effective masses were obtained from the second derivative of the CB energy with respect to wave vector along various
directions away from the valleys minima.
For the VB, the bands are extremely non-parabolic, and so the effective masses could not be evaluated by the method described above.
Instead, we used the six-level \kp\ Dresselhauss-Kip-Kittel model~\cite{ref:13} that depends on three Luttinger parameters, the values of which have been fitted using a conjugate-gradient optimization. 
This optimization is based on a least square error between the curvature masses along the [001], [111] and [110] directions 
obtained with this 6-level \kp\ model and with EPM, \kp\ or GW models. The present extractions technique of the curvature masses and 
Luttinger parameters, although different to the one proposed in Ref.~\onlinecite{ref:24}, gives similar
results (e.g., the Chelikowsky and Cohen-based 
EPM values obtained~in Ge $\gamma_1$=9.563, $\gamma_2$=2.77, $\gamma_3$=3.91 can be compared to our results listed in Table~\ref{tab:table4}). 
The theoretical effective masses and Luttinger parameters are listed in Table~\ref{tab:table4} and compared to experimental data. 
All methods provide reasonably good agreement with experimental values. One should mentioned nevertheless that the present \kp\ model as well as Chelikowsky-based non-local EPM give rather disappointing results for the VB Luttinger 
parameters in Ge. Even if the accuracy of the Luttinger parameters is improved with the present \kp\ model, it still underestimates their values by about 20$\%$. When one expresses the Luttinger parameters in terms of matrix elements~\cite{ref:13}, it becomes clear why it so: 
They contain the term $P/E_{\Gamma_{2'{^l}}}$, the value of which is relatively small in the present work~\cite{ref:non_local}.

\begin{table}  
\caption{\label{tab:table4} Effective curvature masses and Luttinger parameters.}
\begin{ruledtabular}
\begin{tabular}{lcccc}       
Si & Exp. & G$_0$W$_0$& EPM& \kp$^d$\\
 $m_l^{\Delta}$&0.9163$^a$&0.925&0.89&0.928\\
 $m_t^{\Delta}$&0.1905$^a$& 0.189&0.198&0.192\\   
 $m_l^L$&& 1.8083&1.855&1.704\\
 $m_t^L$&&0.1235&0.1535&0.131\\ 
 $\gamma_1$&4.26\footnote{as presented in Ref.~\onlinecite{ref:11}; $^b$I.
 Balslev and P. Lawaetz, as presented in Ref.~\onlinecite{ref:Humphreys};
 $^c$Ref.~\onlinecite{ref:lawaetz}; $^d$present model;
 $^e$Ref.~\onlinecite{ref:Makarov}; $^f$Ref.~\onlinecite{ref:Levinger};
 $^g$Ref.~\onlinecite{ref:Halpern}; $^h$Ref.~\onlinecite{ref:13}. }; 4.285$^a$ &{4.54}&{4.01}&{4.338}\\ 
  & 4.22$^b$; 4.340$^c$&&&\\   
 $\gamma_2$&0.38$^a$; 0.339$^a$ &{0.33}&{0.38}&{0.3468}\\ 
  &0.39$^b$; 0.31$^c$&&&\\
 $\gamma_3$&1.56$^a$; 1.446$^a$&{1.54}&{1.401}&{1.4451}\\      
  &1.44$^b$; 1.46$^c$&&&\\  
\hline 
Ge & Exp.& G$_0$W$_0$ & EPM & \kp$^d$\\
    
$m_l^{\Delta}$&&0.881&0.964&0.874 \\
$m_t^{\Delta}$&&0.176&0.205&0.200\\  
$m_l^L$&1.588$^f$; 1.74$^g$&1.626&1.763&1.59\\
$m_t^L$&0.08152$^f$; 0.079$^g$&0.074&0.099&0.099 \\ 
$\gamma_1$&13.0$^h$; 12.8$^e$;&{13.54}&9.54&10.41 \\ 
&13.25$^a$&\\     
$\gamma_2$&4.4$^h$; 4.0$^e$;&{4.32}&2.75&3.045 \\ 
&4.20$^a$&\\
$\gamma_3$&5.3$^h$; 5.5$^e$;&{5.77}&3.93&4.313 \\      
&5.56$^a$&\\
\end{tabular}   
\end{ruledtabular}           
\end{table}

\begin{figure}
\includegraphics[scale=1.5]{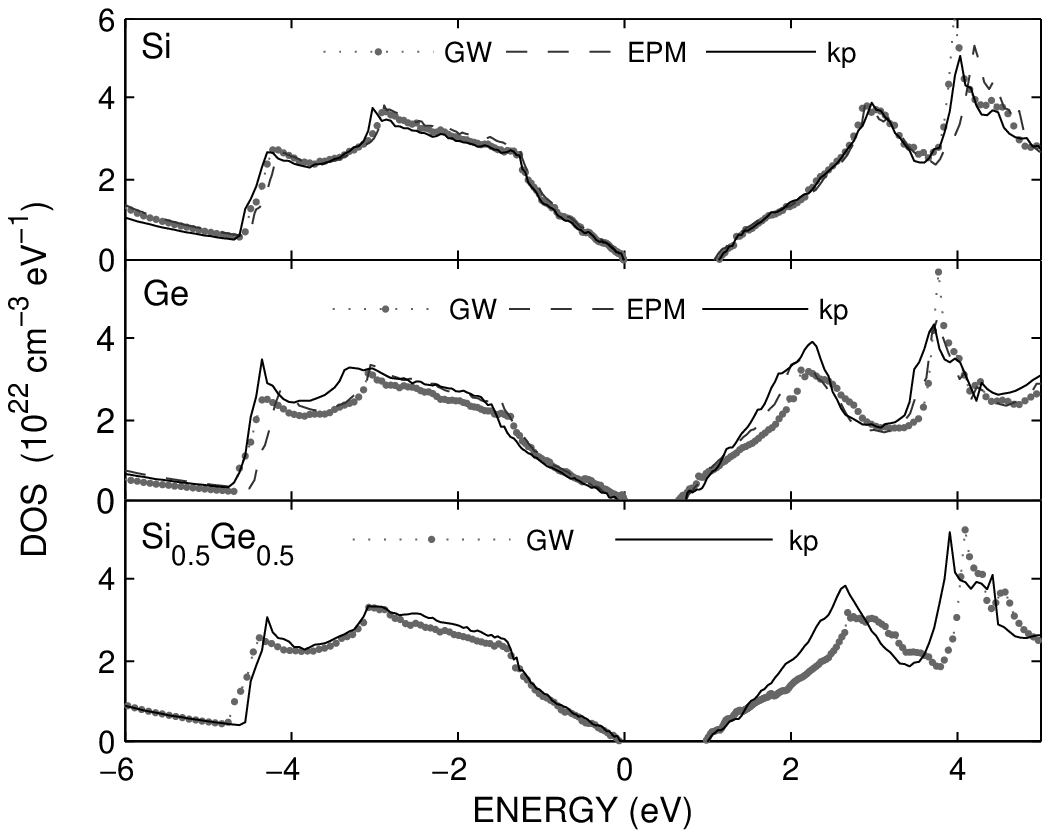}  
\caption{\label{fig:DOS} The DOS of Si, Ge and Si$_{0.5}$Ge$_{0.5}$ alloy: Comparison between thirty-level \kp\ (solid lines), EPM (dashed lines) and GW (dotted lines).}
\end{figure}

Accurate description of the DOS are key features for accurate carrier density
modeling and realistic transport models in semiconductors~\cite{ref:fantini,ref:feraille}. They are also good checks for the quality of the present \kp\ model. Indeed, the DOS not only depends on the band energies but also on their gradient with respect to wave vector (group velocity).
The DOS is obtained using the Gilat and Raubenheimer 
procedure~\cite{ref:16}. We applied exactly the same algorithm using 
the \kp\ (solid lines) the EPM (dashed lines) and the GW models (dotted lines), respectively. As can be seen in Fig.~\ref{fig:DOS},
 the agreement between models is excellent for the VB but also for the CB from -5 eV up to 5 eV.

\subsection{Energy band shifts in strained Si$_{1-x}$Ge$_x$/Si$_{1-y}$Ge$_y$ systems}         
               
\begin{figure}
\includegraphics[scale=1.4]{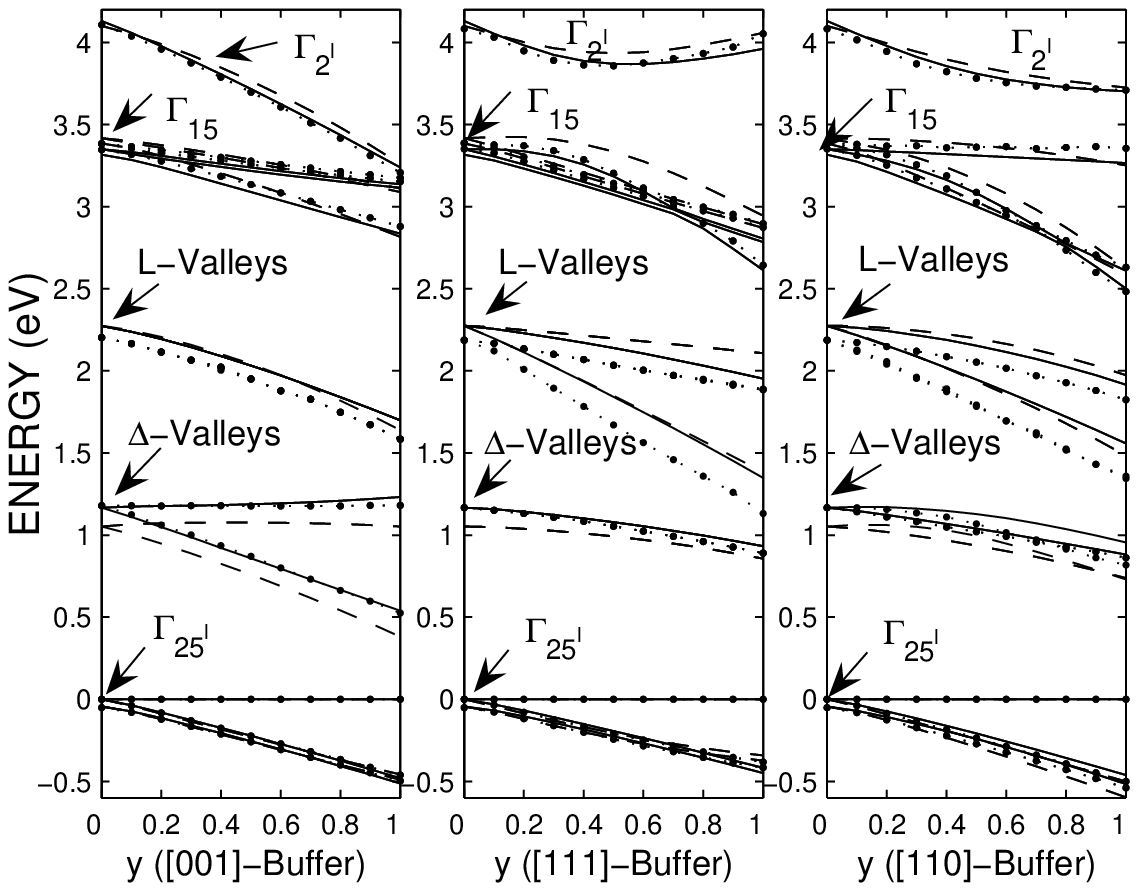}  
\caption{\label{fig:shift_si} Calculated VB and CB shifts of a strained Si layer as a function of $y$-content in the Si$_{1-y}$Ge$_y$ buffer: \kp\ (solid lines), EPM (dashed lines) and GW (dotted lines) simulations performed for various buffer orientations.}
\end{figure}

\begin{figure}
\includegraphics[scale=1.4]{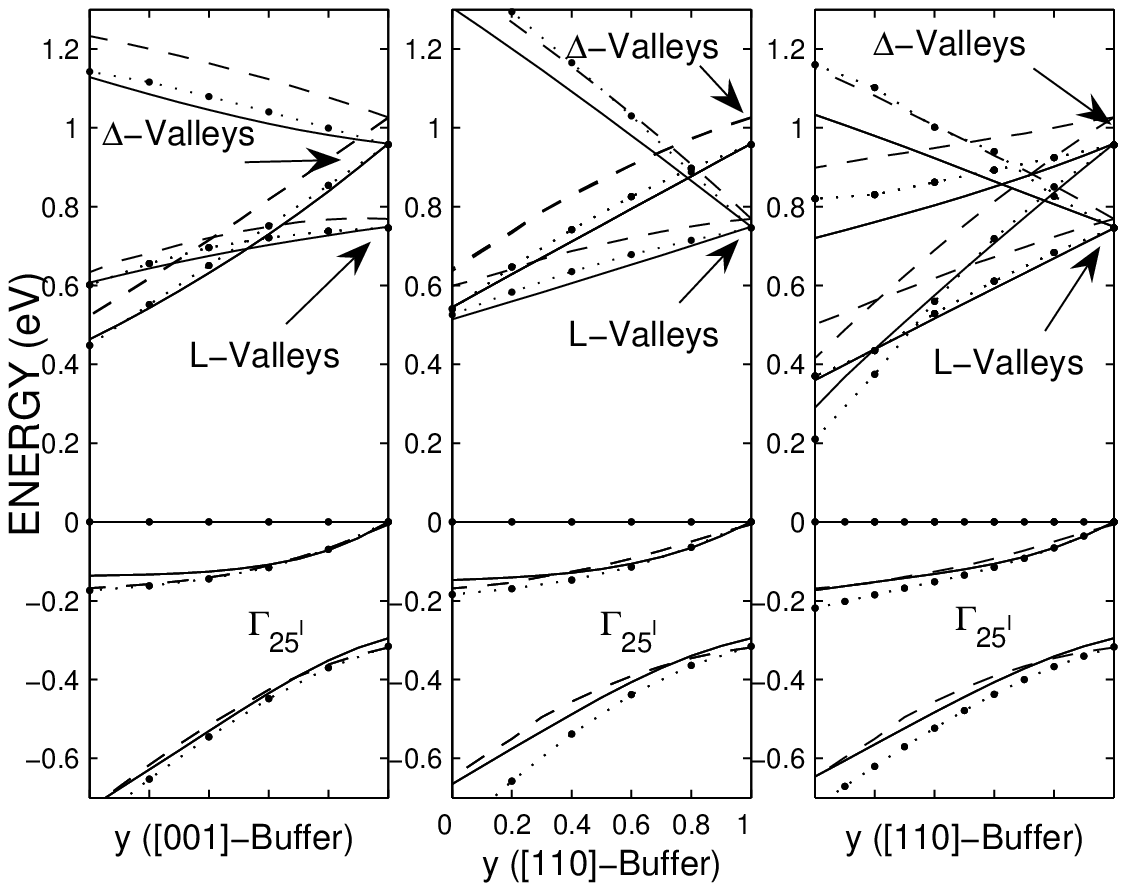}   
\caption{\label{fig:shift_ge} Calculated VB and CB shifts of a strained Ge layer as a function of $y$-content in the Si$_{1-y}$Ge$_y$ buffer: \kp\ (solid lines), EPM (dashed lines) and GW (dotted lines) simulations performed for various buffer orientations.}
\end{figure}

\begin{figure}
 \includegraphics[scale=1.47]{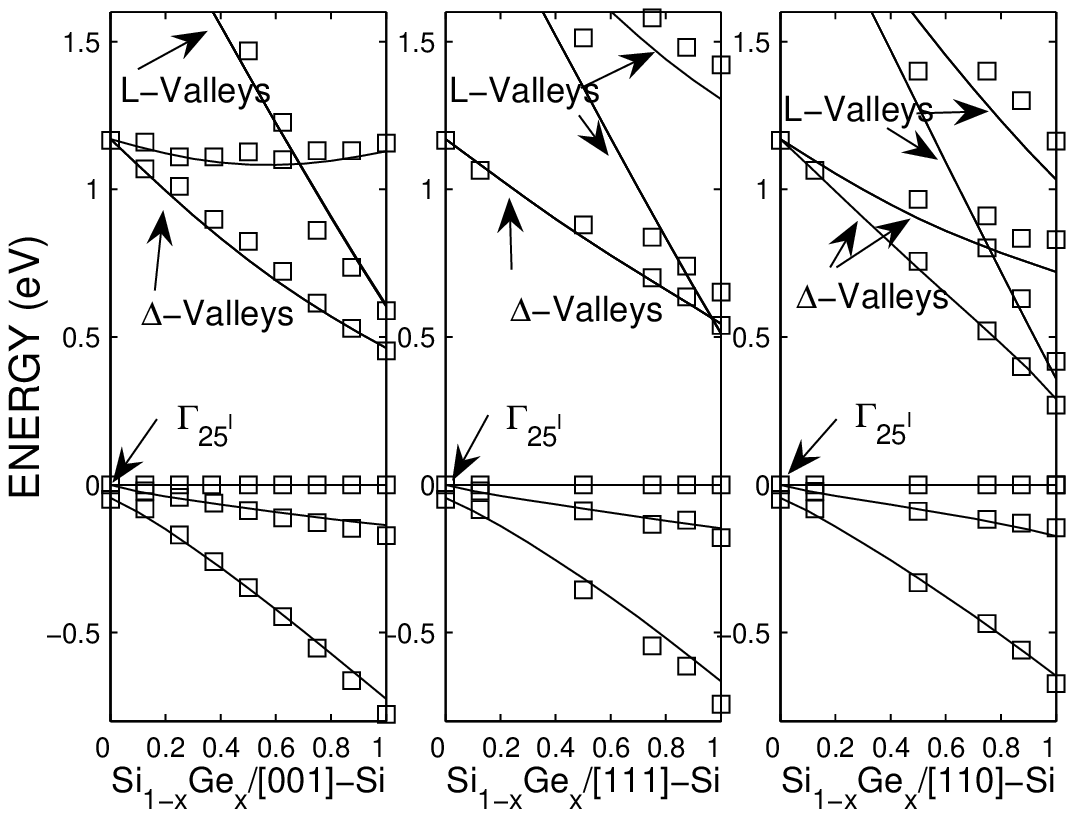}   
\caption{\label{fig:shift_sige} Calculated VB and CB shifts of a strained Si$_{1-x}$Ge$_{x}$ layer grown on a Si buffer: \kp\ (solid lines) and GW (symbols) simulations performed for various buffer orientations.}
\end{figure}

The calculated energy shifts of the main VB and CB extrema in Si, Ge and Si$_{1-y}$Ge$_{y}$ are shown in Fig.~\ref{fig:shift_si},~\ref{fig:shift_ge} and~\ref{fig:shift_sige} for material grown on [001], [111] and [110] Si$_{1-y}$Ge$_y$ buffers 
(the energy scale has been fixed by setting arbitrarily to zero the top of the VBs). 
In these figures, comparison is shown between first principal simulations, EPM, the present \kp\ models and the close agreement between methods can be noticed. In particular, the $\Delta$-valley and L-valley band shifts are correctly modeled with the present \kp\ analysis. 
    

An important effect of strain can be seen for shear distortions ([111] and [110] growth cases) in Figs.~\ref{fig:shift_si}. Several $\Gamma-$eigenvalues shifts exhibit 
a non-linear relation {\it vs.} y. This can typically be inferred from symmetry lowering due to strain. Indeed, several $k$-independent coupling terms ($W_{\Gamma}$ terms in the 
$W_{30\times30}$ perturbation matrix) occur in strained materials (e.g., 
between the $\Gamma_{15}$ and the $\Gamma_{2'^{l}}$ states originally 
non-coupled in the relaxed \kp\ Hamiltonian).
               
From the energy-band shifts, we have calculated the deformation potentials which reflect the 
variation of an individual band energy as a function of applied strain. 
Following the notation of C. G. Van Walle~\cite{ref:Walle}, the experimental values  found in strained Si and Ge for the 
splitting of the top of the VBs and for the lowest CB minima shifts are quoted in Table~\ref{tab:potdef}. The deformation potentials $a_{v}$, $b_{v}$ and $d_{v}$ refer to the hydrostatic, the  
splitting and the shear deformation potentials of the VBs at $\Gamma$, respectively. 
Similar expressions apply for the CB,
 but only $a_c$ is experimentally addressed so far in Si  and Ge~\cite{ref:Cargill}.
Theoretical deformation potentials are consistent with experimental ones excepted for $d_{v}$ in Ge that is overestimated 
by $\simeq$ 8$\%$. This result can be inferred from the slight overestimation of the theoretical internal parameter $\xi$ shown in Table~\ref{tab:elastic}.  
         
Besides deformation potentials at $\Gamma$, other deformation potentials have been experimentally determined at the lowest CB minima, typically along the L-direction for Ge 
and X-direction for Si. In  Table~\ref{tab:potdef}, $\Xi_d^{\Delta}$, $\Xi_d^{L} $ are the hydrostatic CB deformation potentials, 
while $\Xi _u^L$, $\Xi _u^\Delta$  are the splitting deformation potential.   
In general, individual quantities are difficult to measure because they are referred to an absolute scale. 
The case of the VB hydrostatic deformation potentials a$_v$ is worth mentioning: Accurate value is hard 
to obtain from experiments or from theory. In this work, we did not 
calculate a$_v$, we fixed its value following Fischetti and Laux results~\cite{ref:17}: $a_v$=2 for Si, $a_v$=2.1 for Ge and  $a_v=2+0.1x$ for Si$_{1-x}$Ge$_x$.

\subsection{Carrier masses in strained Si$_{1-x}$Ge$_x$/Si$_{1-y}$Ge$_y$ systems}         
     
Changes in effective masses in case of biaxially strained Si and Ge have been been reported by Rieger and Vogl~\cite{ref:24}, Gell~\cite{ref:gell} and by Fischetti and Laux~\cite{ref:17} using EPM. 
In this section, we generalize EPM results to shear distorted crystals and to \kp\ and GW methods. This is to our knowledge the first ab initio 
calculations of the effective masses {\it vs.} strain reported so far in strained Si$_{1-x}$Ge$_{x}$/Si$_{1-y}$Ge$_{y}$ systems. 
                   
\begin{figure}
\includegraphics[scale=1.35]{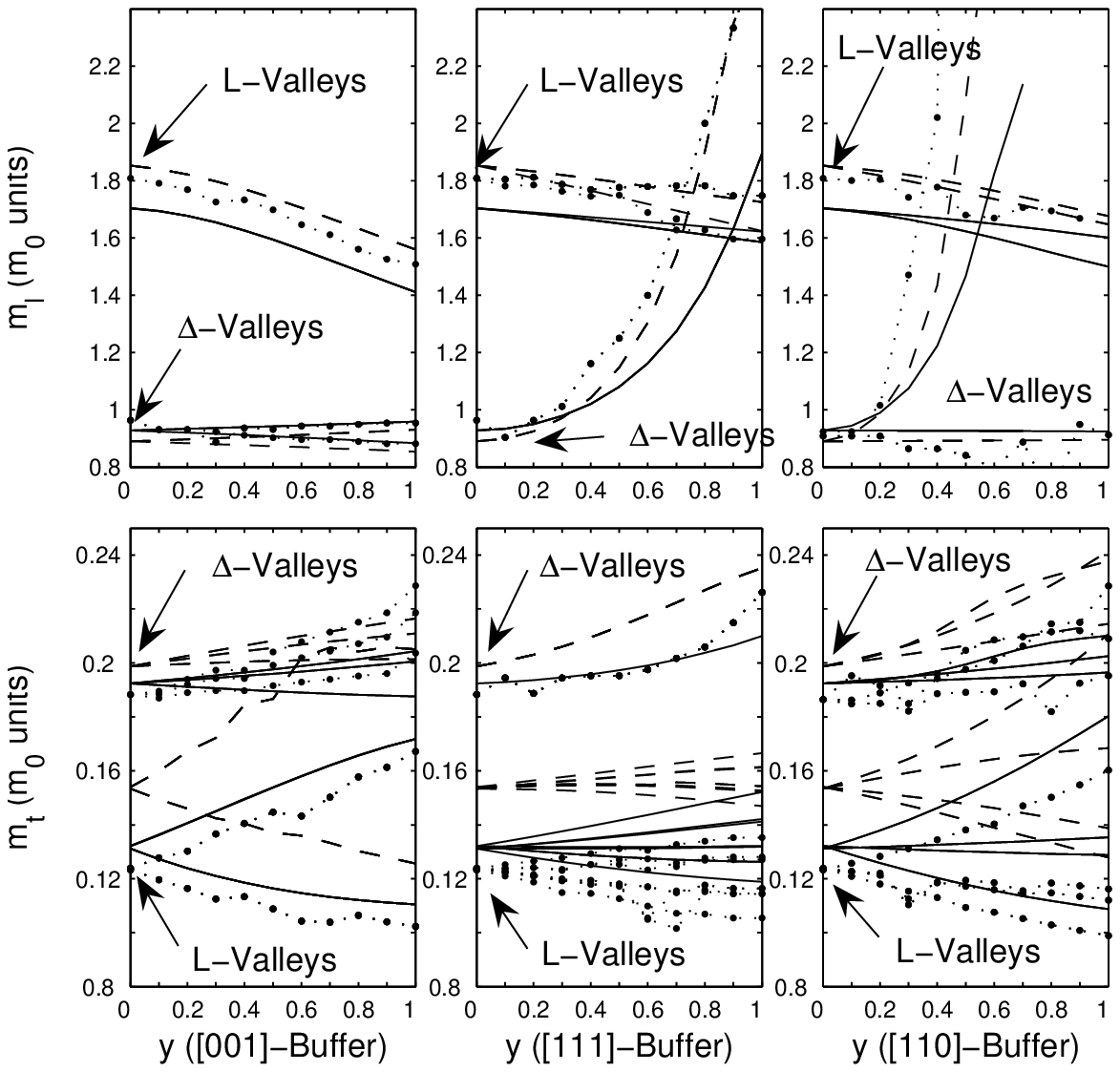}   
\caption{\label{fig:mass_si} Strained Si electron longitudinal and transverse curvature masses as a function of y-content in the Si$_{1-y}$Ge$_y$ buffer: \kp\ (solid lines), EPM (dashed lines) and GW (dotted lines) simulations performed for various buffer orientations.}
\end{figure}

\begin{figure}
\includegraphics[scale=1.35]{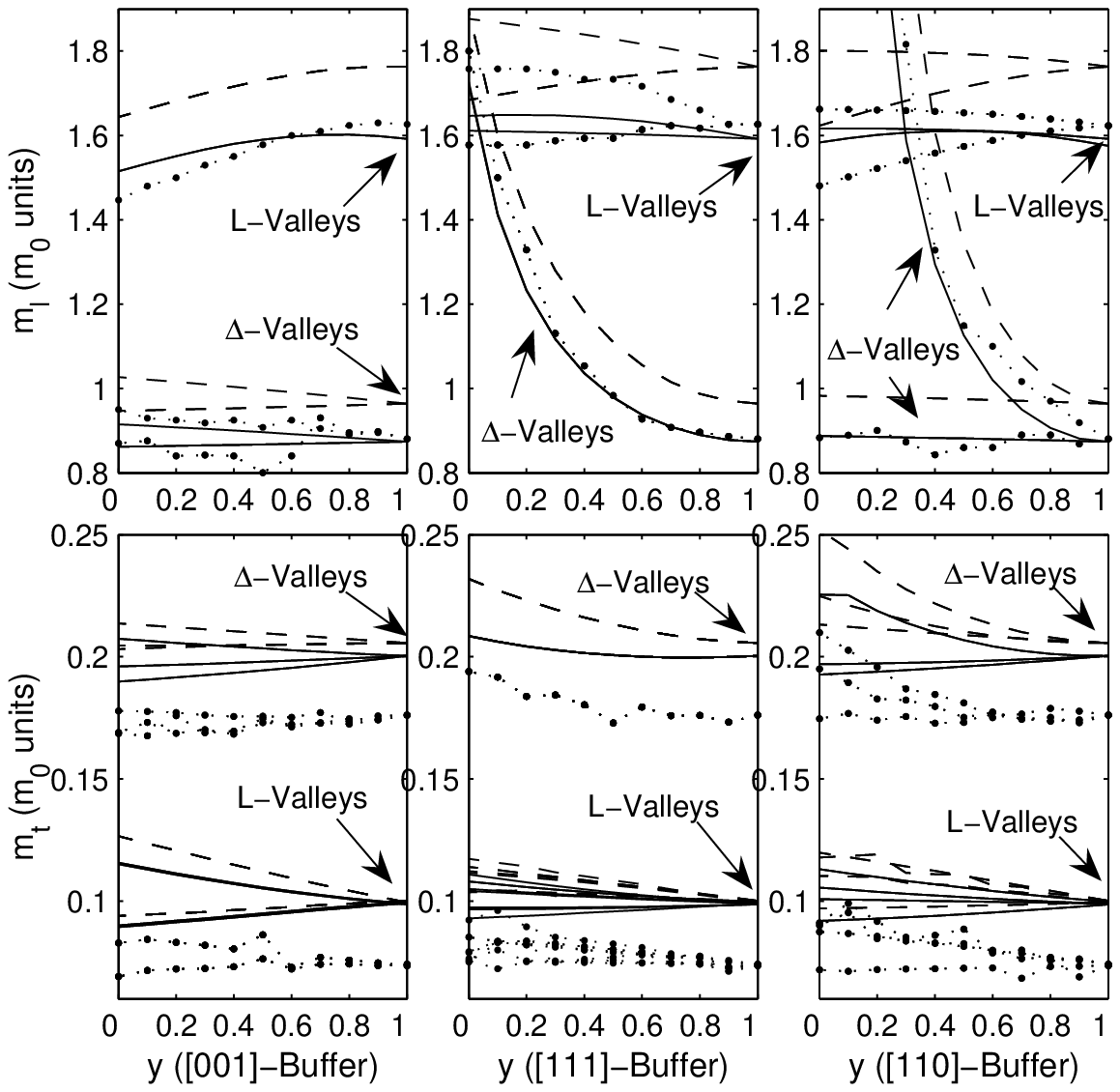}   
\caption{\label{fig:mass_ge} Strained Ge electron longitudinal and transverse curvature masses as a function of y-content in the Si$_{1-y}$Ge$_y$ buffer: \kp\ (solid lines), EPM (dashed lines) and GW (dotted lines) simulations performed for various buffer orientations.}
\end{figure}

Figures~\ref{fig:mass_si} and~\ref{fig:mass_ge} show the strained Si and Ge electron curvature masses as a function 
 of y-content in the Si$_{1-y}$Ge$_y$ buffer for various orientations. According
 to previously reported EPM results~\cite{ref:gell,ref:17,ref:24}, 
electron curvature masses change with applied strain (up to 100$\%$). Simulations performed using semiempirical methods (\kp\, EPM)
 are consistent with first principal results, even though \kp\ tends generally to underestimate the changes in masses {\it vs.} strain.


For biaxially strained Si, the longitudinal and transverse curvature masses depend on the $\Delta$-valley (Fig.~\ref{fig:mass_si}). For the [111] and the [110] growth cases, the curvature masses significantly increase with applied strain. The effects of strain
 on the L-valley curvature masses are generally less pronounced than for the $\Delta$-valley, although variations up to 20$\%$ can be observed.       
Similar conclusions apply for the $\Delta$-valleys and the L-valleys of Ge (Fig.~\ref{fig:mass_ge}).
            
\begin{figure}
\includegraphics[scale=0.4]{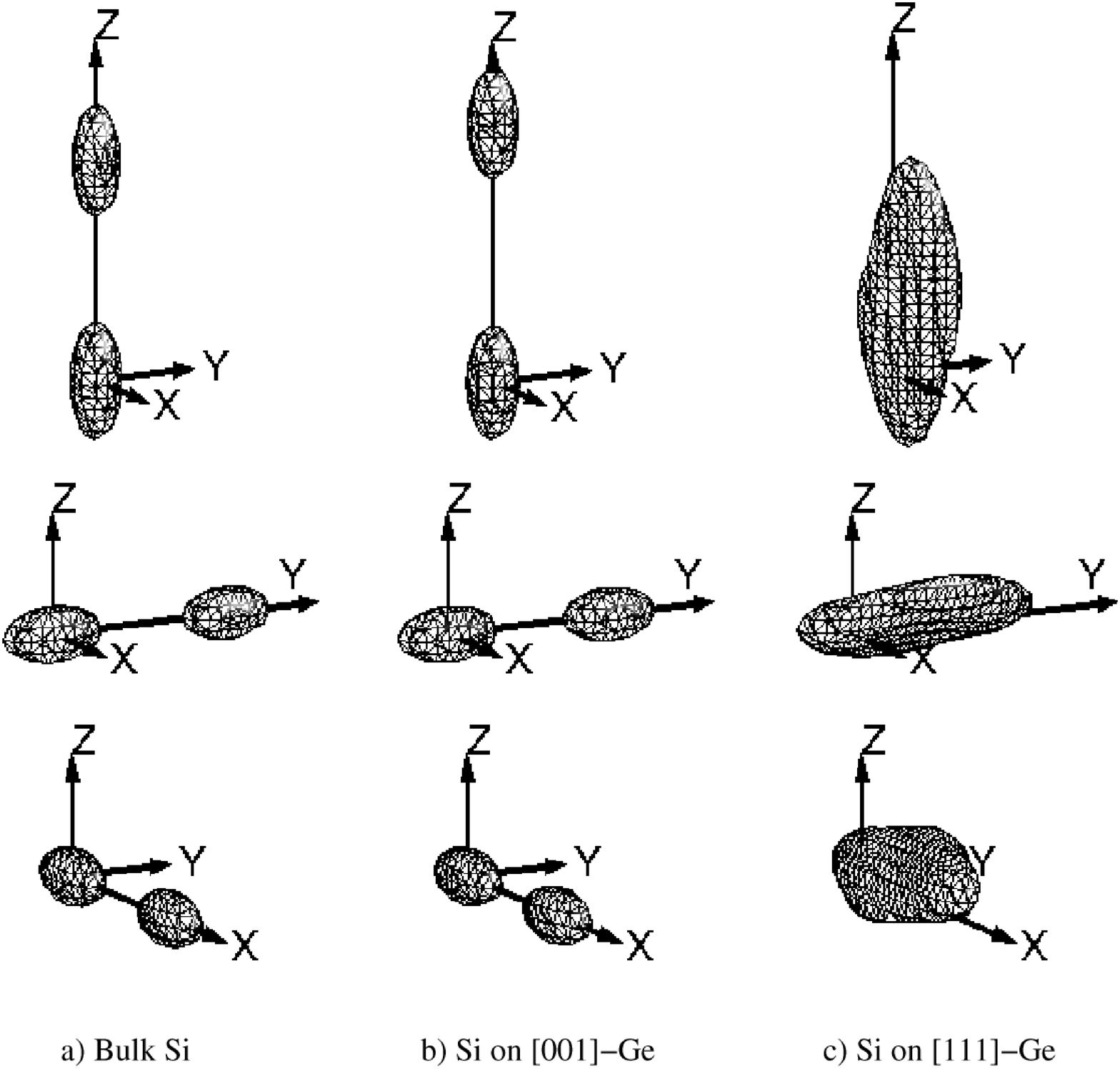}  
\caption{\label{fig:cond3D} 3D-surface plot at thermal energy of the lowest CB across the Brillouin zone edges along the [100], [010] and [001] directions. Simulations performed for a) bulk Si, b) strained Si on [001]-oriented cubic Ge buffer, and c) strained Si on [111]-oriented Ge buffer.}
\end{figure}

We found out that the $\Delta$-valley minima positions in  
reciprocal space also change with applied strain. This behavior is generally more pronounced when shear distortions are applied.
For instance, the $\Delta$-valley minimum distance along $\Gamma$-X direction changes from 84$\%$ in
bulk Si up to 97$\%$ in strained Si on [111]-oriented Ge buffer. This latter point and the changes
in shape can be seen in the three-dimensional (3D) surface plot at thermal energy ($\frac{3}{2}$ kT) shown in Fig.~\ref{fig:cond3D}. 
The $\Delta$-valleys in the first and in the second Brillouin zones are shown respectively along the $\Gamma$-X, $\Gamma$-Y and $\Gamma$-Z directions for bulk Si and for the [100] and the [111]-growth cases. The $\Delta$-valleys 3D-surface in the first and in the second Brillouin zones are clearly separated in the bulk and in the biaxially strained crystal,
 while they "merge" into a single 3D-surface for the [111]-strained crystal.   
      
\begin{figure}
\includegraphics[scale=1.35]{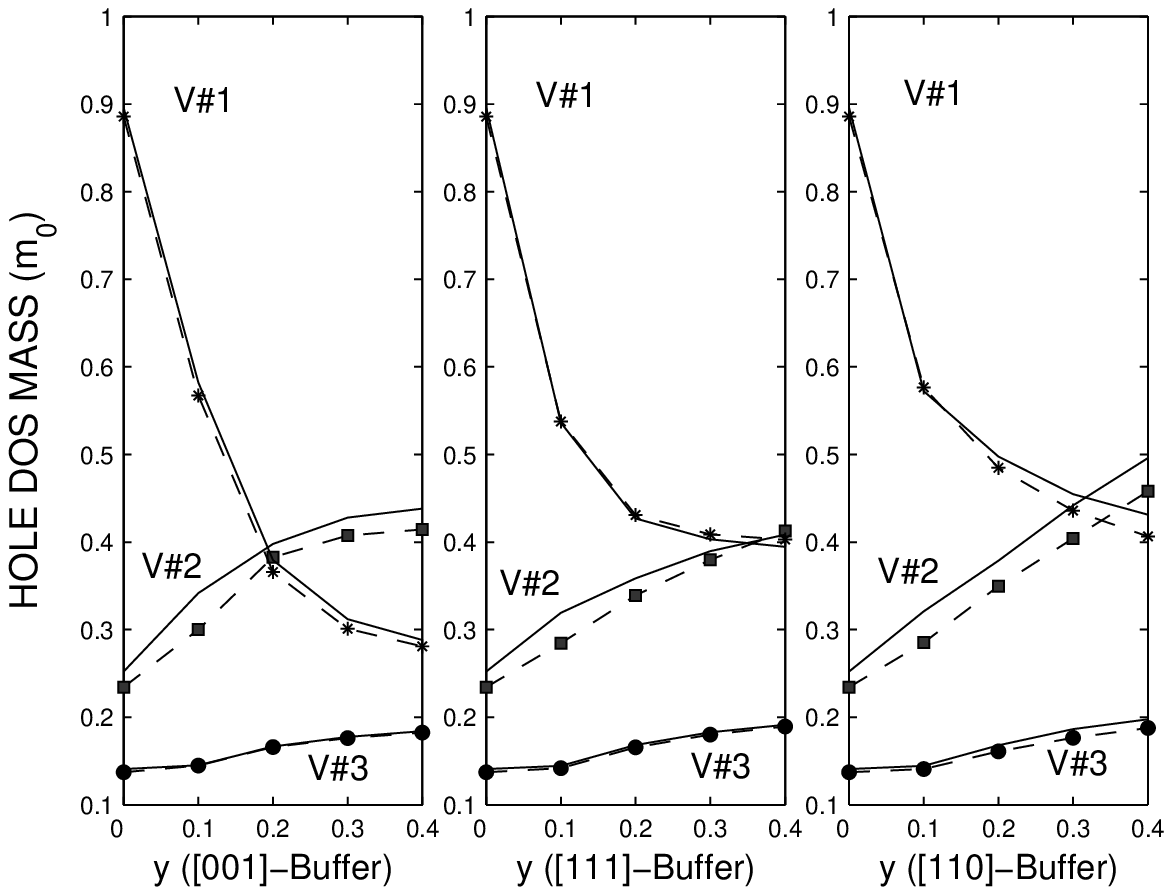}  
\caption{\label{fig:mass_bv_si} Strained Si hole DOS effective masses as a function of y-content in Si$_{1-y}$Ge$_y$ buffer: \kp\ (solid lines) and GW (symbols); T=300 K.}
\end{figure}

\begin{figure}
\includegraphics[scale=1.35]{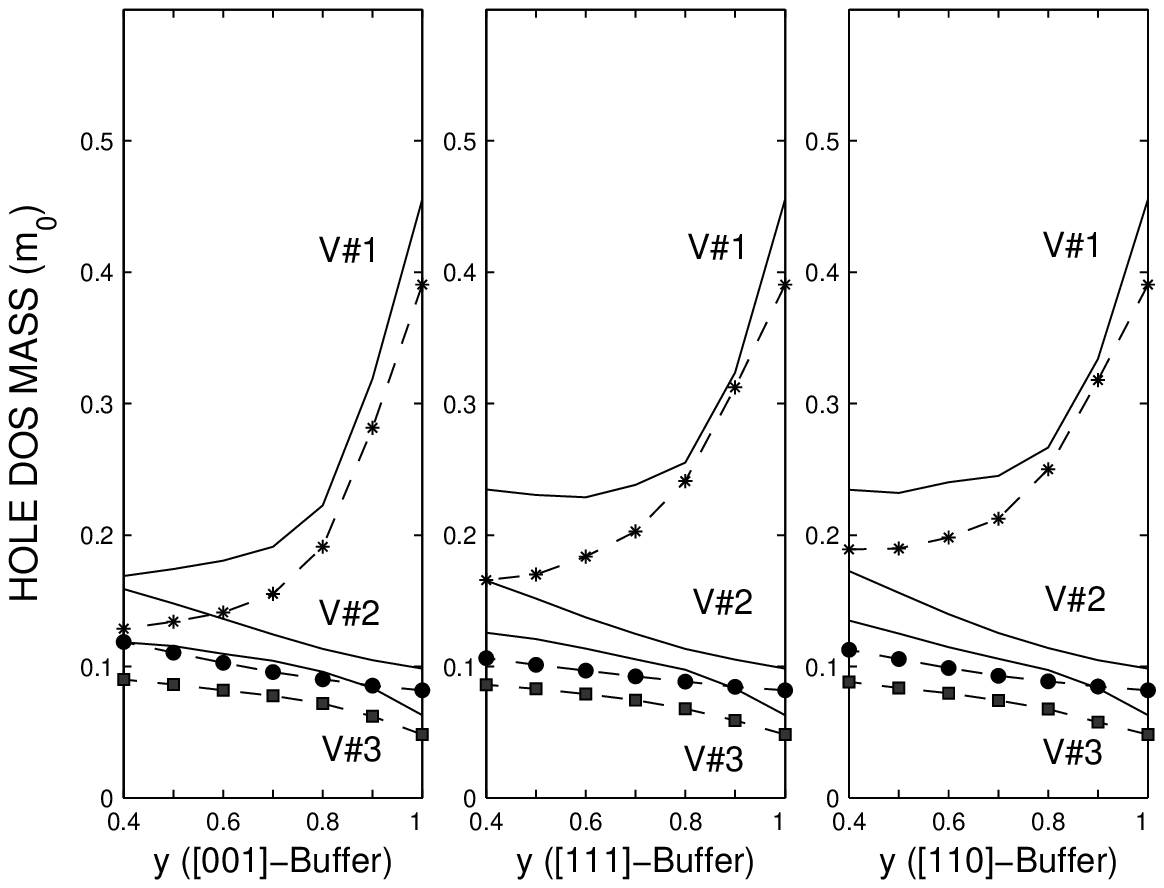}   
\caption{\label{fig:mass_bv_ge} Strained Ge hole DOS effective masses as a function of y-content in Si$_{1-y}$Ge$_y$ buffer: \kp\ (solid lines) and GW (symbols); T=300 K.}
\end{figure}

For the reason mentioned previously, the VB curvature masses cannot be easily evaluated from parabolic fit. Instead, we calculated the DOS 
effective masses at thermal energy. Figures~\ref{fig:mass_bv_si} and~\ref{fig:mass_bv_ge} show the Si and Ge hole DOS effective masses at 300~K as a function of y-content in the buffer. In the unstrained crystal, the heavy-hole DOS mass is approximately three times larger than the light-hole mass. This difference decreases as soon as the degeneracy at
 $\Gamma$ between the heavy and the light hole is removed. One notes that the results shown in Figs.~\ref{fig:mass_si},~\ref{fig:mass_ge},~\ref{fig:mass_bv_si} and~\ref{fig:mass_bv_ge} are consistent with Fischetti's EPM results~\cite{ref:17}.

\section{\label{general}Discussion}      
We have presented a thirty-level \kp\ model for bulk and strained SiGe, the parameters of which have been optimized using a conjugate-gradient procedure on a reference set of energy bands obtained with first principle calculations. The first principle simulations have
been validated through an accurate comparison with experimental published results. 
For bulk SiGe, a set of comparisons with experimental data has shown good agreements 
for the main band gaps values (Table~\ref{tab:table3} and Table~\ref{tab:highsym}), the carrier effective masses and Luttinger parameters (Table~\ref{tab:table4}).
For strained materials, we have benchmarked our first principle results using the deformation potential 
theory applied to the specific case of Si and Ge epitaxial layers grown on [001], [111] and [110]-oriented relaxed buffers. 
Theoretical deformation potentials (Table~\ref{tab:potdef}) have been found to be consistent with experimental ones, within the experimental error. 
A second set of comparisons has been performed using the widely used Chelikowsky and Cohen non-local EPM~\cite{ref:4}, including relativistic 
corrections.        
            
As it is presented in this paper, we obtained good agreements between the present \kp\ model and first 
principle simulations. Since the \kp\ parameters have been fitted on first principle data, one might describe 
these excellent results as a puppy chasing its tail. This is to some degree true, but our procedure yields 
additional pieces of information:
     
(i) Using first principle simulations, important quantities not addressed experimentally, such as high energy levels and 
effective masses at the second CB minima have been taken into account in the present \kp\ model.

(ii) SiGe compounds have been modeled using interpolation functions between Si and Ge \kp\ parameters. 
Additional SO splittings and energy degeneracy removal (due to centrosymmetry breaking) have been accounted for with supplementary coupling \kp\ parameters fitted on first principle data. 

(iii)  State of the art
 "full-band" Monte-Carlo simulators for transport properties in strained Si/Ge
 devices are notably based on EPM~\cite{ref:17,ref:fantini,ref:feraille}.
Because simulations with the present \kp\ model basically give the same results than non-local EPM, the \kp\ model can 
be used as is into a EPM-based MC simulator. It should be noted that the computational burden of the \kp\ 
calculations is impressively reduced in comparison to EPM simulations: The CPU time to obtain a complete band structure is approximatively two 
orders of magnitude less than the CPU time needed by EPM~\cite{ref:transcendental}. 
Further comparisons along that 
line are presented elsewhere~\cite{ref:feraille}.

(iv) Concerning the effective electron masses, our calculations have confirmed
Fischetti and Laux non-local EPM results~\cite{ref:17,ref:24} which have shown that the curvature masses appear to be strongly dependent on the strain. Although never measured 
directly, recent transport simulations in inversion layers using "full-band" MC  
\cite{ref:Irie} have suggested that effective mass change due to strain should be involved in the mobility variation in strained Si and Ge. 
For the effective masses at the second CB minima in energy (i.e.,~L-valley in Si and $\Delta$-valley in Ge), similar good agreements between EPM and GW results have been 
found. It should be noted that these valleys 
could play a significant role in the transport properties of strained semiconductors, particularly for Ge, in which 
$\Delta$-valley and L-valley are separated by only 200 meV. It has been shown in Section III, that biaxially strained Ge on Si$_{1-y}$Ge$_y$ 
buffer exhibits a cross-over between $\Delta$-valley and L-valley minima for $y \leq$ 0.5. 
         
For the reasons mentioned above, the present highly optimized \kp\ parameters set improves the published \kp\ parameters set of Refs.~\cite{ref:9} and~\cite{ref:richard} for bulk materials and extends its predictions to SiGe alloy. 
Furthermore, the Pikus and Bir perturbative treatment of strain was for the first time evaluated in the Cardona and Pollak thirty-level \kp\ formalism~\cite{ref:9}. We have shown notably that this correction captures 
the main feature of strained-crystal band structures, such as energy shifts versus strain and effective masses change due to strain. 
One should mention that this is not the case with the recently published twenty-level \kp\ model for biaxially strained Si and Ge~\cite{ref:Richard1}, in which 
important contributions to the strain perturbation matrix have been omitted. We found out that the
behavior of the CBs in strained semiconductors strongly depend on the W$_k$ and W$_{\Gamma,\Gamma}$ terms in the perturbation matrix. Neglecting these
 terms, as in Ref.~\onlinecite{ref:Richard1}, leads to a large underestimation of the CB equienergy valley splitting and effective masses changes versus strain. Moreover, VB time reversal symmetry at X
 and correct band-shifts at L-valleys versus strain cannot be obtained without these contributions.

\section{\label{Conclusion}Conclusion}  
In this paper, we have developed a highly optimized thirty-level \kp\ model for strained Si, Ge and SiGe alloys.
          
A series of ab initio DFT-LDA simulations that include GW correction and relativistic effects in Si, Ge ans SiGe alloys has been performed 
with a view to obtaining informations not addressed by experiments.
Once a reference set of energy bands has been obtained, we have optimized the \kp\ model parameters using a conjugate-gradient procedure in order to fit as close as possible 
first principle results, but also carrier effective masses and luttinger parameters. A simple interpolation between Si and Ge \kp\ parameters 
has been proposed in order to model SiGe alloy. 
  
The electronic structure of strained Si$_{1-x}$Ge$_{x}$/Si$_{1-y}$Ge$_{y}$ systems has been studied using 
first principle simulations. For the first time, the well-known Pikus and Bir correction~\cite{ref:bir} for strained materials has been examined 
within this thirty-level \kp\ formalism. The deformation potentials have been obtained 
from first principle simulations in order to fit the shifts of the thirty lowest energy levels at $\Gamma$ {\it vs.} applied strain, but also the general shape of 
the band structure of the strained crystal. 
    
Finally, the present \kp\ model has been validated through an accurate set of comparisons with experimental data in relaxed and strained Si, Ge, and Si$_{1-x}$Ge$_{x}$ alloys.
 A second set of comparisons with first-principal simulations, but also with the widely used Chelikowsly and 
Cohen non-local EPM \cite{ref:4} has also shown a good agreement. 
The present \kp\ description of strained Si, Ge and SiGe accurately reproduces
 the overall band structure, as 
well as the band shifts, the carrier effective masses and the DOS {\it vs.} applied strain.

\begin{acknowledgments} 
We would like to thank Y. M. Niquet for a number of useful discussions and N. Kauffman
for a critical reading of the manuscript.
\end{acknowledgments}

\newpage

\appendix

\begin{widetext}
\section{\kps\ matrix for relaxed materials}
 
The thirty-level \kp\ matrix (Eq. \ref{eq:2}) for relaxed materials writes: 
      
\begin{equation}
H^{30}_{k.p}=\left[ 
\begin{array}{cccccccccccccccc}
H_{\Gamma_{2'^u}}^{2\times2} && P''' H_k^{2\times6}  && 0 && 0 &&  0 && 0  && 0 && P'' H_k^{2\times6}\\
\vspace{.15 cm}   
 && H_{\Gamma_{25'^u}}^{6\times6} && R' H_k^{6\times4}  && 0 && 0 && Q' H_k^{6\times6}&& P' H_k^{6\times2} && H^{so}_{\Gamma_{25'^u},\Gamma_{25'^l}}\\
 \vspace{.15 cm}  
&&  && H_{\Gamma_{12'}}^{4\times4} && 0 && 0 &&  0 && 0 && R H_k^{4\times6}  \\
\vspace{.15 cm}
 &&  &&  &&H_{\Gamma_{1^u}}^{2\times2}  && 0 && T H_k^{2\times6} && 0 && 0\\
\vspace{.15 cm}
 &&  &&  &&  &&H_{\Gamma_{1^l}}^{2\times2}  && T' H_k^{2\times6} && 0 && 0\\
\vspace{.15 cm}
 &&  &&   &&  &&  &&H_{\Gamma_{15}}^{6\times6}  && 0 && Q H_k^{6\times6} \\
\vspace{.15 cm}
 &&  &&  &&  &&  &&  && H_{\Gamma_{2'^l}}^{2\times2} && P H_k^{2\times6} \\  
\vspace{.15 cm}
 &&  &&  &&  &&  &&  &&  && H_{\Gamma_{25'^l}}^{6\times6}
\end{array}        
\right],
\label{eq:H3030}
\end{equation}
whose diagonal blocks read:
\begin{equation}
\left\{ 
\begin{array}{ll} 
H^{6\times6}_{\Gamma}=&diag \left( E_{\Gamma}+\frac{\hbar^2k^2}{2m} \right) +H^{SO}_{\Gamma}\\
H^{4\times4}_{\Gamma}=&diag \left( E_{\Gamma}+\frac{\hbar^2k^2}{2m} \right)\\
H^{2\times2}_{\Gamma}=&diag \left( E_{\Gamma}+\frac{\hbar^2k^2}{2m} \right) \\
   
\end{array}
\right.,
\label{eq:diag}
\\
\end{equation}   
where $k^2=k^2_x+k^2_y+k^2_z$, and $diag()$ stands for the diagonal matrix. $E_\Gamma$ is the eigenvalue of the state labeled by $\Gamma$, as listed in Table~\ref{tab:table3}. 
The coupling constants ($P,P',P'''$, etc ...) are listed in Table~\ref{tab:table2}. 
$H^{SO}_{\Gamma}$ is the SO matrix, which depends on the SO coupling parameters listed in Table~\ref{tab:table2}:  
\begin{equation}    
H^{SO}_{\Gamma}=\frac{\Delta_{\Gamma}}{3} \left[   
\begin{array}{cccccc}
 -1 & -i& 0 & 0 & 0 & 1\\ 
 i & -1 & 0 & 0 & 0 & -i\\ 
 0 & 0 & -1 & -1 & i & 0\\
 0 & 0 & -1 & -1 & i & 0\\ 
 0 & 0 & -i & -i & -1 & 0\\
 1 & i & 0 & 0 & 0 & -1\\
\end{array}
\right].
\label{eq:HSO}
\end{equation}
The non-zero \kp\ blocks write:

\begin{equation}
\left\{ 
\begin{array}{ll}
H_{k}^{6\times6}=&\left[ \begin{array}{cccccc}
 0 & k_z & k_y & 0 & 0 & 0\\
 k_z & 0 & k_x & 0 & 0 & 0\\
 k_y & k_x & 0 & 0 & 0 & 0\\
 0 & 0 & 0 & 0 & k_z & k_y\\
 0 & 0 & 0 & k_z & 0 & k_x\\
 0 & 0 & 0 & k_y & k_x & 0\\
\end{array}
\right]
\vspace{.5 cm}
\\ 
     
H_{k}^{4\times6}=&\left[ 
\begin{array}{cccccc}
 0 & \sqrt{3}k_y & -\sqrt{3}k_z & 0 & 0 & 0\\
 2k_x& -k_y &  -k_z  &0& 0 & 0\\
 0 & 0 & 0  & 0 & \sqrt{3}k_y  & -\sqrt{3}k_z\\
0  & 0 & 0& 2k_x& -k_y &  -k_z\\
\end{array}
\right]

\vspace{.5 cm}
\\

H_{k}^{2\times6}=&\left[ 
\begin{array}{cccccc}
 k_x & k_y & k_z & 0 & 0 & 0 \\ 
 0 & 0 & 0 & k_x & k_y & k_z \\      
\end{array}
\right]
 
\end{array}
\right..   
\label{eq:H6-6}
\end{equation}
       
We have finally included a SO coupling term 
between the $\Gamma_{25^u}$ and the $\Gamma_{25^l}$ states and fitted the coupling strength $\Delta_{\Gamma,\Gamma}$ in order to respect the time reversal degeneracy at X.

\section{\kps\ matrix for strained materials}
The perturbation matrix for strained materials (Eq.~\ref{eq:strcor}) writes:
\begin{equation}     
W^{30}_{k.p}=\left[ 
\begin{array}{cccccccccccccccc}
W_{\Gamma_{2'^u}}^{2\times2} && P''' W_k^{2\times6}  && W_{\Gamma_{2'^u},\Gamma_{12'}}^{2\times4} && 0 &&  0 && W^{2\times6}_{\Gamma_{2'^u},\Gamma_{15}}&& W^{2\times2}_{\Gamma_{2'^u},\Gamma_{2'^l}} && P'' W_k^{2\times6}\\
\vspace{.15 cm}
 && W_{\Gamma_{25'^u}}^{6\times6} && R' W_k^{6\times4}  && W^{6\times2}_{\Gamma_{25'^u},\Gamma_{1^u}} && W^{6\times2}_{\Gamma_{25'^u},\Gamma_{1^l}} && Q' W_k^{6\times6} && P' W_k^{6\times2} && W_{\Gamma_{25'^l},\Gamma_{25'^u}}^{6\times6} \\
\vspace{.15 cm}
 &&  && W_{\Gamma_{12'}}^{4\times4} && 0 && 0 &&  W_{\Gamma_{12'},\Gamma_{15}}^{4\times6} && W^{4\times2}_{\Gamma_{12'},\Gamma_{2'^l}}  && R W_k^{4\times6}  \\ 
\vspace{.15 cm}  
 &&  &&  &&W_{\Gamma_{1^u}}^{2\times2}  && W^{2\times2}_{\Gamma_{1^u},\Gamma_{1^l}} && T W_k^{2\times6} && 0 && W^{2\times6}_{\Gamma_{1^u},\Gamma_{25'^l}}\\     
\vspace{.15 cm}    
 &&  &&  &&  &&W_{\Gamma_{1^l}}^{2\times2}  && T' W_k^{2\times6} && 0 && W^{2\times6}_{\Gamma_{1^l},\Gamma_{{25'}^l}} \\       
\vspace{.15 cm}
 &&  &&   &&  &&  &&W_{\Gamma_{15}}^{6\times6}  && W^{6\times2}_{\Gamma_{15},\Gamma_{{2'}^l}}  && Q W_k^{6\times6} \\   
\vspace{.15 cm}
 &&  &&  &&  &&  &&  && W_{\Gamma_{2'^l}}^{2\times2} && P W_k^{2\times6} \\           
\vspace{.15 cm} 
 &&  &&  &&  &&  &&  &&  && W_{\Gamma_{25'^l}}^{6\times6}   
\end{array}
\right].      
\label{eq:W3030}     
\end{equation}
           
There are two types of coupling terms in the matrix described by Eq.~\ref{eq:W3030}; k-independent terms (labeled $W_{\Gamma}$) coming from the second term in Eq.~\ref{eq:strcor}
and terms that are linear in $k$ (labeled $W_k$) coming from the first term in Eq. \ref{eq:strcor}.
         
The k-independent $W_{\Gamma}$ blocks write:
\begin{equation}
\left\{ 
\begin{array}{ll}
W_{\Gamma}^{6\times6}=&\left[ 
\begin{array}{cc}
W_{\Gamma}^{3\times3}& 0\\
 0 & W_{\Gamma}^{3\times3}\\   
\end{array}      
\right]    
\vspace{.5 cm}  
\\

W_{\Gamma}^{3\times3}=&\left[ 
\begin{array}{cccccc}
l\epsilon_{xx}+m{\left(\epsilon_{yy}+\epsilon_{zz}\right)}& n\epsilon_{xy}& n\epsilon_{xz}\\
  n\epsilon_{xy} & l\epsilon_{yy}+m{\left(\epsilon_{xx}+\epsilon_{zz}\right)}& n\epsilon_{yz} \\   
 n\epsilon_{xz}& n\epsilon_{yz} &  l\epsilon_{zz}+m{\left(\epsilon_{xx}+\epsilon_{yy}\right)}\
\end{array}      
\right]     
   
\vspace{.5 cm}
\\

W^{4\times4}_{\Gamma_{12}} =&\left[ 
\begin{array}{cccc}
 A{\epsilon}_{xx}+B\left({\epsilon}_{yy}+{\epsilon}_{zz}\right) & E\left({\epsilon}_{yy}-{\epsilon}_{zz}\right)& 0  & 0\\
 E\left({\epsilon}_{yy}-{\epsilon}_{zz}\right) & C{\epsilon}_{xx}+D\left({\epsilon}_{yy}+{\epsilon}_{zz}\right) & 0 &  0\\
 0  & 0 & A{\epsilon}_{xx}+B\left({\epsilon}_{yy}+{\epsilon}_{zz}\right)& E\left({\epsilon}_{yy}-{\epsilon}_{zz}\right)\\    
 0  & 0& E\left({\epsilon}_{yy}-{\epsilon}_{zz}\right) & C{\epsilon}_{xx}+D\left({\epsilon}_{yy}+{\epsilon}_{zz}\right)\\    
\end{array}   
\right]   

\vspace{.5 cm}
\\

W^{2\times2}_{\Gamma} =a_{\Gamma}&\sum_i\left[    
\begin{array}{cc}
{\epsilon}_{ii} & 0\\
0 & {\epsilon}_{ii}\\
\end{array}    
\right]

\vspace{.5 cm}
\\

W^{2\times6}_{\Gamma} =f_{\Gamma} &\left[      
\begin{array}{cccccc}
 \epsilon_{yz}&\epsilon_{xz}&\epsilon_{xy}&0& 0 &0\\
0& 0&0 &\epsilon_{yz}&\epsilon_{xz}&\epsilon_{xy}\\        
\end{array}  
\right]

\vspace{.5 cm}
\\   

W^{4\times2}_{\Gamma} =g_{\Gamma}& \left[    
\begin{array}{cccccc}
\sqrt{3}\left(\epsilon_{yy}-\epsilon_{zz}\right)& 0 \\       
2\epsilon_{xx}-\epsilon_{yy}-\epsilon_{zz} & 0 \\ 
0&\sqrt{3}\left(\epsilon_{yy}-\epsilon_{zz}\right)\\ 
0 & 2\epsilon_{xx}-\epsilon_{yy}-\epsilon_{zz}\\  
\end{array}     
\right]
       
\vspace{.5 cm}
\\   
 
W^{4\times6}_{\Gamma} =h_{\Gamma}& \left[    
\begin{array}{cccccc}
0 & \sqrt{3}\epsilon_{zx} & -\sqrt{3}\epsilon_{xy}&0&0&0 \\       
2\epsilon_{yz}& -\epsilon_{zx} &  -\epsilon_{xy}&0&0&0 \\       
0&0&0&0 & \sqrt{3}\epsilon_{zx} & -\sqrt{3}\epsilon_{xy} \\       
0&0&0&2\epsilon_{yz}& -\epsilon_{zx} &  -\epsilon_{xy}\\       
\end{array}     
\right]

\end{array}
\right..
\label{eq:W24}    
\end{equation}

The deformation potentials ($l,n,m$...etc) are listed in Table~\ref{tab:coupkp} (the coefficients not mentioned in the Table are set to zero). 
Group theory considerations allow to write the five coefficients $A,B,C,D,E$ as a linear combination of four coefficients:

\begin{equation} 
\left\{ 
\begin{array}{ll}
 A = & 6\left(b_{12}-d_{12}\right)\\
 B = & 3\left(a_{12}+b_{12}-2c_{12}\right)\\
 C = & 2\left(2a_{12}-4c_{12}+b_{12}+d_{12}\right)\\
 D = & 5b_{12}-2c_{12}-4d_{12}+a_{12}\\
 E = & \sqrt{3}\left(2c_{12}-2d_{12}-a_{12}+b_{12}\right)\\  
\end{array}
\right..    
\label{eq:ABC}             
\end{equation}

The first term in Eq. \ref{eq:strcor} gives rise to an additional non-diagonal k-independent 
coupling between states of the same polarity. These blocks and the corresponding deformation potentials have 
been labeled using a double subscript notation (e.g.,~$W_{\Gamma_{12},\Gamma_{2^u}}$). For simplicity, we 
dropped the double subscript notation for the coupling between identical states (e.g.,~$W_{\Gamma_{12}}$).         
     
Finally, the \kp\ coupling terms due to the second term in Eq. \ref{eq:strcor} write:       
 
\begin{equation}  
\left\{ 
\begin{array}{ll}
W^{6\times6}_{k}=&-\sum_i\left[   
\begin{array}{cccccc}
 0 & \epsilon_{iz}{k_{i}}& \epsilon_{yi}{k_{i}}& 0 & 0 & 0\\
 \epsilon_{iz}{k_{i}} & 0 & \epsilon_{xi}{k_{i}} & 0 & 0 & 0\\
 \epsilon_{iy}{k_{i}} & \epsilon_{ix}{k_{i}} & 0 & 0 & 0 & 0 \\
 0 & 0 & 0 & 0 & \epsilon_{iz}{k_{i}} & \epsilon_{yi}{k_{i}}\\
 0 & 0 & 0 &\epsilon_{iz}{k_{i}} & 0 & \epsilon_{xi}{k_{i}}\\
 0 & 0 & 0 &\epsilon_{iy}{k_{i}} & \epsilon_{ix}{k_{i}} & 0\\
\end{array}
\right]
\vspace{.5 cm}
\\   
         
W_{k}^{4\times6}=&-\sum_i\left[ 
\begin{array}{cccccc}
 0 & \sqrt{3}\epsilon_{iy} k_i & -\sqrt{3}\epsilon_{iz} k_i & 0 & 0 & 0\\
 2\epsilon_{ix} k_i& -\epsilon_{iy} k_i &  -\epsilon_{iz} k_i  &0& 0 & 0\\
 0 & 0 & 0  & 0 & \sqrt{3}\epsilon_{iy} k_i  & -\sqrt{3}\epsilon_{iz} k_i\\
0  & 0 & 0& 2\epsilon_{ix} k_i& -\epsilon_{iy} k_i &  -\epsilon_{iz} k_i\\
\end{array}    
\right]
    
\vspace{.5 cm}  
\\   
W_{k}^{2\times6}=&-\sum_i \left[ 
\begin{array}{cccccc}
 \epsilon_{ix} k_i & \epsilon_{iy} k_i & \epsilon_{iz} k_i & 0 & 0 & 0 \\ 
 0 & 0 & 0 & \epsilon_{ix} k_i & \epsilon_{iy} k_i & \epsilon_{iz} k_i \\      
\end{array}
\right]   
 
\end{array}
\right..
\label{eq:W6-6}
\end{equation}
where $i$ stands for $x$, $y$ and $z$.
\end{widetext}


\begin{thebibliography}{xxxxxxx55555}
\bibitem{ref:Star} In the present paper, in order to distinguish the corrected GW data from the original G$_0$W$_0$ results, the subscripts are dropped out whenever the correction is applied.
\bibitem{ref:noteksi}Experimental values for $\xi \in $ [0,1]. However as pointed out by Levine~\cite{ref:levine} most of the theoretical internal strain parameter
 calculation~\cite{ref:nielsen}, and~\cite{ref:wei} have used a non standard definition of $\xi$
 (i.e., $\xi' = \frac{2}{3} \xi$) which implies $\xi' \in$ [0 $\frac{2}{3}$]. In this text we used $\xi'$, and for simplicity the ' is dropped in the notations.
\bibitem{ref:smoothing} For the purpose of obtaining a continous description in K, U, as required for instance in MC simulators~\cite{ref:feraille}, 
we found a satisfactory solution using a smoothing function on $P'''$ parameter reducing locally (in the near-K region) its value to 
1.3 a.u..
\bibitem{ref:normaliz} The present \kp\ matrix uses for $\Gamma_{12}$ the functions $\gamma_{1^{-}}$ and $\gamma_{2^{-}}$ 
as in Ref.~\cite{ref:13}, while Cardona and Pollak~\cite{ref:9} use $\frac{1}{\sqrt{2}}(\gamma_{1^{-}}-\gamma_{2^{-}})$ and $\frac{1}{\sqrt{2}}(\gamma_{1^{-}}+\gamma_{2^{-}})$. 
The coefficients R and R' are $\frac{1}{\sqrt{2}}$ smaller in this work.
\bibitem{ref:non_local} As written in Eq.~(\ref{eq:2}), the \kp\ Hamiltonian does not explicitly
include non-local correction to the purely local operator $V(r)$. In this work, the non-local
effects have been "renormalized" into the \kp\ parameters. In Ge where these effects are
prominent~\cite{ref:4}, certain coefficients (e.g., $P=1.186$ a.u.) are smaller than Cardona and
Pollak~\cite{ref:9} ones (e.g., $P=1.36$ a.u.).
\bibitem{ref:ourepm} The present EPM simulations in strained material are based on the method shown in Ref.~\cite{ref:17}. However in order to improve the accuracy of the deformation potentials and energy band shifts with experimental data, we used a slightly different form-factors interpolation function.
\bibitem{ref:transcendental}In general, for EPM, it is necessary to 
use a set of $100-200$ plane waves. On the contrary, a set of only thirty $\Gamma$-centered Bloch functions
 is sufficient in the \kp\ method. This leads to a substantial difference    
in terms of computation time.
Moreover, in the relativistic non-local EPM~\cite{ref:4} a significant computation effort is spent
evaluating transcendental functions (e.g.,~Bessel functions) while in the \kp\ Hamiltonian only algebraic terms linear in $k$ are calculated.
\end{thebibliography}
\end{document}